\documentclass[fleqn,usenatbib]{mnras}

\usepackage{newtxtext,newtxmath}
\usepackage[T1]{fontenc}

\DeclareRobustCommand{\VAN}[3]{#2}
\let\VANthebibliography\thebibliography
\def\thebibliography{\DeclareRobustCommand{\VAN}[3]{##3}\VANthebibliography}

\usepackage{graphicx}	% Including figure files
\usepackage{amsmath}	% Advanced maths commands
\usepackage{xcolor}
\usepackage{float}
\usepackage{bm}
\title[SBI of the ionization history from 21 cm 2DPS]{Simulation based inference pipeline of the ionization history from the 2D 21 cm power spectrum}

\author[Cooper, Norregaard, et al.]{
Nadia Cooper$^{1}$\thanks{These authors contributed equally.}\thanks{E-mail: nadia.cooper19@imperial.ac.uk}%
,
Carina Norregaard$^{1}$\footnotemark[1] \thanks{E-
mail: c.norregaard22@imperial.ac.uk}%
,
Romain Meriot$^{1}$
, 
Jonathan R. Pritchard$^{1,2}$%
\\
$^{1}$Blackett Laboratory, Imperial College London, Prince Consort Road, London, SW7 2AZ, UK\\
$^{2}$Max-Planck-Institut für Radioastronomie, Auf dem Hügel 69, D-53121 Bonn, Germany
}

\date{Accepted XXX. Received YYY; in original form ZZZ}

\pubyear{\the\year{}}

\begin{document}
\label{firstpage}
\pagerange{\pageref{firstpage}--\pageref{lastpage}}
\maketitle

\begin{abstract} 
The 21 cm signal contains a wealth of information about the formation of the first stars and the reionization of the intergalactic medium during the Cosmic Dawn (CD) and Epoch of Reionization (EoR). The timing of these important milestones has only roughly been constrained through indirect measurements (e.g. cosmic microwave background (CMB) optical depth, and Lyman-$\alpha$ forest). Therefore, inferring the neutral fraction over cosmic time is a goal of upcoming 21 cm experiments, such as the Square Kilometer Array (SKA). We contrast two approaches to infer astrophysical parameters and ionization history from 21 cm 2D power spectra (2DPS). We develop an emulator of the 21 cm 2DPS, trained on 21cmFAST simulations, taking into account the expected instrumental noise from the SKA and sample variance. We then perform simulation based inference (SBI) using neural posterior estimation (NPE). We compare training on datasets of noisy 2DPS obtained from 21cmFAST simulations and the emulator, to infer astrophysical parameters of interest. Using an emulator of the ionization history, we then obtain posterior distributions of the ionization history over the redshift range z $\sim$ 5-12. We demonstrate that both methods are capable of accurately recovering the ionization history and astrophysical parameters. However, coverage tests indicate that using a larger number of emulated samples instead of simulated samples does not improve predictions. This work suggests that due to the stochastic nature of the 2DPS, a more complex architecture than a dense model, with built in stochasticity, is needed to better emulate the 2DPS by accurately capturing the sample variance.
\end{abstract}


%%%%%%%%%%%%%%%%%%%%%%%%%%%%%%%%%%%%%%%%%%%%%%%%%%

%%%%%%%%%%%%%%%%% BODY OF PAPER %%%%%%%%%%%%%%%%%%

\section{Introduction} \label{sec: intro}

The 21 cm transition, arising from the spin-flip emission line of neutral hydrogen, is a promising probe of the early universe. The signal can be seen in many important phases of cosmic history. The Cosmic Dawn (CD) saw the formation of the first stars and galaxies, which reionized the neutral hydrogen in the intergalactic medium (IGM) during the Epoch of Reionization (EoR). Due to the high abundance of hydrogen, the 21 cm signal provides a tracer of the properties of the gas. This signal contains a wealth of information which can be used to constrain the properties of the IGM, providing an insight into the first stars and galaxies.
\linebreak
\linebreak
Many radio telescopes currently have the EoR 21 cm signal as a target. These include GMRT \citep{paciga2013simulation}, LOFAR \citep{mertens2020improved, ghara2020constraining, ceccotti2025first}, HERA \citep{deboer2017hydrogen}, and MWA \citep{barry2019improving, li2019first, kolopanis2023new}. These experiments have provided upper limits on the 21 cm power spectra \citep{ mertens2020improved, trott2020deep, abdurashidova2022first}. However, observing this signal remains an observational challenge. Models of the 21 cm signal estimate its magnitude to be $\sim 10^{4}$ times fainter than the foregrounds \citep{chapman2019foregrounds}. While the removal of foregrounds remains a challenge, several methods are commonly used to mitigate their effect, such as an independent component analysis method like FastICA \citep{Hyvarinen, 2012MNRAS.423.2518C}, or Gaussian Process Regression (GPR) \citep{2018MNRAS.478.3640M}.  Furthermore, works such as \cite{2021MNRAS.500.2264H, 2024mertensforegrounds} demonstrate the successful recovery of the cosmological signal through machine learning methods.

The Square Kilometre Array (SKA) \citep{koopmans2015cosmic} is predicted to make a detection of the 2D power spectra (2DPS), the cylindrically averaged power spectra, and potentially map the full topography of the signal. The difficulties in modeling and subtracting foreground emissions motivated the SKA Science Data Challenge 3a \citep{bonaldi2025square}, where participants were asked to remove foregrounds, instrumental and systematic effects, which concluded with promising results. The possibility of an upcoming detection in the coming years has prompted renewed interest how best to infer the reionization history from the 2D power spectrum. This is the subject of the SKA Science Data Challenge 3b (Bonaldi et al. in prep). Previous works include \cite{2016MNRAS.457.1864L}, who presented a framework for inferring ionization history from 1DPS, providing the first forecasts of HERA's ability to constrain the ionization history.

Indirect observations have begun to constrain the timing of important milestones in the early universe, such as the cosmic microwave background (CMB) optical depth $\tau$ \citep{aghanim2020planck, de2021inference, heinrich2021reionization}, the Lyman-$\alpha$ forest \citep{fan2006constraining, becker2007evolution, bosman2018new, d2023xqr}, damping wings in quasar spectra \citep{bolton2011neutral, mortlock2011luminous, banados2018800, wang2020significantly, yang2020poniua}, and Lyman-$\alpha$ emission from galaxies \citep{ouchi2010statistics, clement2012evolution, konno2014accelerated, drake2017muse, hoag2019constraining, shibuya2019morphologies}. 

To perform Bayesian inference on a sufficient set of data, semi-numerical codes can be used to reduce computation time - e.g. $\textsc{SimFast21}$ \citep{Simfast21Santos}, $\textsc{21cmFAST}$ \citep{mesinger201121cmfast}, $\textsc{21cmSPACE}$ \citep{Visbal2012, Fialkov2012}, $\textsc{GRIZZLY}$ \citep{GRIZZLYGhara}, $\textsc{BEoRN}$ \citep{Schaeffer_2023}. The simulation code $\textsc{21cmFAST}$, developed by \cite{mesinger201121cmfast}, for example uses excursion-set formalism and perturbation theory to create 3D density, spin temperature, and 21-cm brightness temperature fields \citep{mesinger201121cmfast}. However, these semi-numerical codes can still be computationally costly, and so neural network models known as emulators have grown in popularity \citep{ Kern_2017, Schmit_2017, Jennings2018, ghara2020constraining, Bye_2022, Breitman_2023, meriot2025comparison}. These emulators are trained on a suite of simulations to mimic the behavior of the simulation code itself, maintaining accuracy and reducing computation time. \cite{Schmit_2017}, \cite{ghara2020constraining}, and \cite{Breitman_2023} emulated the 1D power spectrum from the astrophysical parameters, and showed the emulators' application in a Bayesian inference framework. 21cmEMU, the emulator created by \cite{Breitman_2023}, is trained on 1.28M samples generated from 21cmFAST. They output several summary statistics, including the neutral fraction, 21cm brightness temperature, and the spherically averaged power spectrum \citep{Breitman_2023}. Emulators like 21cmEMU can reduce the time for one evaluation from 1 hour, in the case of $\textsc{21cmFAST}$, to 0.1 s.  \cite{Bye_2022} develop an emulator called 21cmVAE, which specifically emulates the global 21cm signal from seven astrophysical parameters. These mentioned works have focused on the 1D power spectra and have all successfully used their emulators in inference frameworks. Since the 2DPS contains more information \citep{prelogovic2024informative}, we focus on building a 2DPS emulator with the purpose of testing its efficacy in an inference pipeline. Applying 2D power spectrum emulators in the field of 21 cm cosmology has yet to be explored. We assess whether emulating the 2D power spectrum as opposed to the 1DPS is a useful technique for parameter inference.

The goal of these emulators is to obtain constraints via inference. Markov chain Monte Carlo (MCMC) using the 21 cm power spectra is a common approach to inference from the 21 cm signal \citep[e.g.][]{greig201521cmmc}. However, this requires a functional form of the likelihood, often assumed Gaussian, which may not be accurate for 21 cm inference. MCMC inferences also require $\sim 10^{5}$ samples to converge. These simulations are required to be generated on the fly, which can make exploratory works expensive. Emulators can help reduce the computational time for MCMC pipelines significantly. In order to circumvent these issues, 21 cm inference using Simulation Based Inference (SBI) has grown in popularity \citep[e.g.][]{zhao2022simulation, saxena2024simulation, greig2024inferring, meriot2025comparison}. Here, the likelihood is implicitly modeled through an amortized bank of samples from the joint distribution. SBI can learn the statistical relationship between data and parameters. In this work we use Neural Posterior Estimation (NPE) to directly model the posterior density over all of prior space.

We perform SBI on noisy 21 cm 2DPS to obtain posteriors on three astrophysical parameters of interest; the virial temperature ($T_{\mathrm{vir}}$), ionizing efficiency ($\zeta$), and photon mean free path, ($R_{\mathrm{mfp}}$). We compare two methods to do this. Firstly, with SBI trained on samples of the 2DPS of 21cmFAST simulations, and secondly, with SBI trained on samples obtained via an emulator of the 2DPS of the 21 cm signal. Since 21 cm simulations are computationally expensive, we are constrained by simulation budget on the sampling density of our training set for the SBI pipeline. The goal of the 2DPS emulator was to increase sampling density of the prior by using the emulator to interpolate between samples in the training set. We also infer the ionization history of the universe between redshifts $z = 5$ and $z = 12$ in 41 redshift bins. We train an emulator to predict the ionization history from astrophysical parameters. We then sample from our learnt posteriors on the astrophysical parameters, and predict the ionization history at these points in parameter space. In this way, we can obtain posterior distributions on the neutral fraction of hydrogen. One of the goals of this research was to determine whether the use of an emulator improved constraints with the SBI, due to increased sampling density. 

This paper is structured as follows, in Section \ref{21cm} we introduce the relevant 21 cm background. In Section \ref{data} we discuss our data model used to generate the training set. The emulators for the 2DPS and ionization history are discussed in Section \ref{emulators}. Section \ref{SBI} gives some background into simulation based inference and neural posterior estimation and the results of our inference are shown. Discussion is given in Section \ref{discussion}, followed by concluding remarks. Appendix \ref{ion_corner} shows the posterior distribution for our recovered neutral fractions. Appendix \ref{architectures} highlights the effects of architecture choice on the inference outcome. 

\section{21 cm signal}\label{21cm}

Here, we summarize a basic introduction into 21 cm physics. For a more detailed review, we encourage the reader to refer to \cite{furlanetto2006cosmology, pritchard201221, liu2020data}.
\linebreak
The detectability of the 21 cm signal in the early universe hinges on deviations between the brightness temperature of neutral hydrogen, and a background radio source, usually the CMB. The brightness temperature of the 21 cm signal $T_{b}$ is determined by the thermal and ionization state of the IGM. It is given by, 
\begin{equation}
    \delta T_b = \frac{T_S-T_{\mathrm{CMB}}}{1+z}(1-e^{-\tau_{21}}),
\end{equation}
where $T_S$ is the spin temperature of the neutral hydrogen, $\tau_{21}$ is the 21 cm optical depth, which can be re-expressed to give the following equation for the brightness temperature \citep{loeb2001reionization}:
\begin{equation} \label{bright_temp}
\begin{aligned}
\delta T_b \approx\; & 27\, x_{\mathrm{HI}}(1+\delta_b)
\left(\frac{\Omega_b h^2}{0.023}\right)
\left(\frac{0.15}{\Omega_m h^2} \cdot \frac{1+z}{10}\right)^{1/2} \\
& \times \left(\frac{T_S - T_R}{T_S}\right)
\left[1 - \frac{1}{H(z)(1+z)} \frac{\partial v_r}{\partial r} \right] \, \text{mK},
\end{aligned}
\end{equation}

where $x_{\mathrm{H I}}$ is the neutral fraction of hydrogen and $\delta_b$ is the fractional overdensity of baryons. The final term is due to the velocity gradient along the line of sight. $T_S$, the spin temperature of neutral hydrogen, is defined via the relation between the number densities, $n_i$, of hydrogen atoms in the two hyperfine levels. This ratio between the levels is given by
\begin{equation}
    n_1/n_0 = (g_1/g_0) e^{(-T_\star/T_S)}.
\end{equation}

Here, $g_1/g_0 = 3$ is the ratio of statistical degeneracy, and $T_\star \equiv hc/k\lambda_{\mathrm{21 cm}} = 0.068 K$. There are three different processes which contribute to the spin temperature. These are the scattering of CMB photons, collisions with other hydrogen atoms and electrons, and Lyman-$\alpha$ photons \citep{wouthuysen1952excitation, field1958excitation}. The equation relating them is \citep{field1958excitation}, 

\begin{equation}
    T^{-1}_{S} = \frac{T_{\gamma}^{-1} +x_{\alpha}T_\alpha^{-1} + x_cT_k^{-1} }{1+x_\alpha + x_c}. 
\end{equation}

Here, $T_\gamma$ is the temperature of the bath of radio photons, set by the CMB ($T_\gamma = T_{\mathrm{CMB}}$). $T_\alpha$ is the temperature of the Lyman-$\alpha$ radiation. $x_c$ and $x_\alpha$ are coupling coefficients of the atomic collisions and the Ly$\alpha$ photons respectively. 

At different phases of cosmic history, different coupling processes dominate. During the dark ages, collisional coupling to the cold gas dominates, causing an absorption feature in the globally averaged signal \citep{loeb2004measuring}. Due to the expansion of the universe, collisional coupling eventually becomes ineffective, and the spin temperature decouples from the gas, hence, $T_{S}$ rises to $T_{\mathrm{CMB}}$. The first galaxies then generate a UV background, exciting via the Wouthuysen-Field effect \citep{wouthuysen1952excitation, field1958excitation, hirata2006wouthuysen}, producing another absorption feature. X-ray photons then heat the IGM. As the spin temperature couples to the hot IGM, we see an emission feature in the 21 cm signal. As the neutral fraction tends to zero, spin-flip emissions cease, and the 21 cm signal tends to zero. In this paper, we work in the saturated regime $T_S \gg T_{\gamma}$.

\section{Simulated dataset}\label{data}
\subsection{21cmFAST simulations}
In order to train our emulator, we required a large dataset, and so opted to use the semi-numerical code 21cmFAST  \citep{mesinger201121cmfast, murray202021cmfast, qin2020tale, munoz2022impact, davies2025efficient}. Semi-numerical codes are computationally less expensive than full radiative transfer codes like C2Ray \citep{mellema2006c2}, Licorice \citep{semelin201721ssd, meriot2025comparison} and  RAMSES–CUDATON \citep{ocvirk2016cosmic, ocvirk2020cosmic, lee2025line}. 21cmFAST creates lightcones in the given redshift range generated with the chosen astrophysical input parameters.  

Our simulated dataset consists of lightcones generated between z $\sim$ 5-12, and the three astrophysical parameters we used to create the simulations were:

\begin{itemize}
    \item log$_{10}T_{\mathrm{vir}}$ (K): the minimum virial temperature required to produce ionizing radiation. In order for stars to form, gas must cool enough to condense. There are several mechanisms for cooling in order for the gas to collapse into stars, one of which is through radiative cooling from molecular hydrogen \citep{Haiman_1997}. This cooling mechanism is effective for `minihalos' before reionization, where the halos are thought to be virialized at $T_{\mathrm{vir}}$ $\sim$ $10^3$ K \citep{Haiman_2000, Shapiro_2006, Nadler_2025}. However, the molecular hydrogen cooling is disrupted by stellar feedback, which dissociates the molecules and stops the cooling process \citep{Haiman_1997}. Efficient star formation then appears in halos with a corresponding $T_{\mathrm{vir}}$ $\sim$ $10^4$ K, where atomic hydrogen cooling takes over \citep{Haiman_2000, Barkana_2002}. For our analysis, we let $\mathrm{log}_{10} T_{\mathrm{vir}}$ (K) vary from $3.3$ - $5$, to encompass all star-forming halos masses.

    \item $\zeta$: ionizing efficiency of star-forming galaxies. The ionization efficiency is calculated as follows: 
    \begin{equation}
        \zeta = 30\left(\frac{f_{\mathrm{esc}}}{0.15}\right)\left(\frac{f_{\star}}{0.1}\right)\left(\frac{N_{\gamma}}{4000}\right)\left(\frac{2}{1+n_{\mathrm{rec}}}\right),
    \end{equation}
    where $f_{\mathrm{esc}}$ is the fraction of UV photons that escape, $f_{\star}$ is the fraction of collapsed gas in stars, $N_{\gamma}$ is the number of ionizing photons per baryon produced, and $n_{\mathrm{rec}}$ is the average number of recombinations of a hydrogen atom in the IGM \citep{furlanetto2006cosmology, SobacchiMesinger2014}. In this work, the 21cmFAST simulations were created varying $\zeta$, not its constituent parts. The expected value for $N_{\gamma}$ is $\sim$ 4000 for Pop II stars \citep{Barkana_2001}, while the expected values for $f_{\mathrm{esc}}$ and $f_{\star}$ lie around 0.1 for both \citep{Robertson_2015}. We vary $\zeta$ over the range $10 \leq \zeta \leq 50$.

    \item R$_{\mathrm{mfp}}$ (Mpc): mean free path of ionizing photons through HII regions. Star forming regions produce ionizing photons that go on to escape their HII region and ionize neutral HI. However, these photons encounter Lyman limit systems (LLS), which are dense neutral hydrogen pockets where the recombination rate is higher than in the IGM \citep{Miralda_Escude_2000}. The ionizing photons are absorbed in these regions, limiting their ionizing radius. The mean free path of these photons has been measured $\sim$ 50 Mpc at $z =  6$ \citep{Prochaska2009}. However, \cite{Mesinger_2012kSZ} find that when the ionized region radius exceeds that of the mean free path of the ionizing photons, the process of reionization is impeded, implying that the smaller R$_{\mathrm{mfp}}$ values can have a large effect.  We therefore vary the $R_{\mathrm{mfp}}$ between 3-80 Mpc to match the work of \citep{Mesinger_2012kSZ}.
\end{itemize}

We sampled from a Latin Hypercube (LH) of uniform prior across the ranges given above for each parameter, creating 39,000 samples.  For each parameter combination, we generate a lightcone using 21cmFAST, each with a different seed dictating the underlying density distribution initialized at z = 300 \citep{mesinger201121cmfast}. We used a box size of 250 Mpc with 128 cells per side and eight logarithmic $k_{\perp}$ and $k_{\parallel}$ bins between 0.01-3.01 and 0.02-2.05, respectively. To mimic expected observations of the EoR, we subtract the mean temperature from the lightcone, we then separate the lightcone into 15 MHz chunks, creating a total of eight chunks, and calculate the 2D power spectrum of each. The central redshifts associated with these eight frequency intervals are: 11.22, 9.82, 8.70, 7.80, 7.05, 6.45, 5.88, 5.41. This results in eight 2D power spectra for each parameter sample.  The emulator model is then trained on 27,300 samples, validated on 7,800, and tested on 3,900 samples. The test, train, validate split is kept consistent throughout the pipeline; in training the 2DPS emulator, the ionization history emulator and the SBI framework. This ensures any observation on which we have performed inference has not been seen by any step in the pipeline. The same is true for the test set used to assess coverage in \ref{SBC}.

\subsection{Noise model}\label{sec:noise}

Noise was generated using the Tools21cm package \citep{giri2020tools21cm}. UV maps were constructed from the 2016 Vogel SKA antenna configuration (consisting of 512 antenna) with a box size of 250 Mpc, a total observation time of 1,000 hrs, observation time per day of four hours with an integration time of ten seconds, pointed at a declination of -30.00 Dec.

Tools21cm generates noise according to the following model. The noise is dependent on the system temperature according to the following equation:
\begin{equation}
    \Phi_{\mathrm{system}} = \frac{2 k_{B} T_{\mathrm{Sys}}}{A_{\mathrm{eff}}},
\end{equation}
where $\Phi_{\mathrm{system}}$ is the equivalent flux density of the radio antenna, $k_{B}$ is the Boltzmann constant, $T_{\mathrm{Sys}}$ is the system temperature, and $A_{\mathrm{eff}}$ is the effective area of the antenna. The system temperature can be further broken down into,
\begin{equation}
    T_{\mathrm{Sys}} = T_{\mathrm{sky}} + T_{\mathrm{receiver}},
\end{equation}

where, $T_{\mathrm{sky}}$ scales with frequency as:
\begin{equation}\label{frequency noise scaling}
    T_{\mathrm{sky}} = 60 \left(\frac{300}{\nu}\right)^{2.55},
\end{equation}

as determined by the foregrounds. This frequency dependence on the noise means that the observed power spectrum at higher redshift has a lower SNR. 

The standard deviation of the noise map is given by:

\begin{equation}\label{eq:noise dist}
\sigma =\mathrm{RMS}_{\mathrm{noise}} \times \sqrt{\frac{2 t_{\mathrm{int}}}{t_{\mathrm{obs}}\  N_{\mathrm{ant}}(N_{\mathrm{ant}}-1)}}, 
\end{equation} 
where $\mathrm{RMS}_{\mathrm{noise}}$ is the root mean square of the noise per visibility,
\begin{equation}
    \mathrm{RMS}_{\mathrm{noise}} = \frac{\Phi_{\mathrm{system}}}{\sqrt{2 \Delta\nu t_{\mathrm{int}}}},
\end{equation}

and $t_{\mathrm{int}}$ is the integration time, $t_{\mathrm{obs}}$ is the observation time, $N_{\mathrm{ant}}$ is the number of antenna, and $\Delta\nu$ is the frequency depth. Tools21cm generates noise lightcones distributed according to Equation \ref{eq:noise dist}.

We created lightcones of noise that matched the specifications of our 21cmFAST simulations, and calculated the power spectra of the noise in the same 15 MHz frequency chunks.  

We generated 10,000 noise realizations using Tools21cm. From this we calculated the mean power spectra of the noise. In order to reduce computational cost when obtaining noise realizations, we sampled from the mean power spectra of the noise with the following distribution \citep{mcquinn2006cosmological}:

\begin{equation}\label{noise_dist_equation}
\delta P_{\mathrm{Noise}}(\textbf{k},z) =\mathcal{N}\left(\mu=0, \sigma^2 = \frac{2 \hat{P}_{\mathrm{Noise}}^{2}(\textbf{k},z)}{N_{\mathrm{modes}}(\textbf{k},z)}\right).
\end{equation}

Here, $\hat{P}_{\mathrm{Noise}}(\textbf{k},z)$ is the mean power spectra of the noise, and $N_{\mathrm{modes}}$ is the number of modes per pixel.

Furthermore, the database generated for this work assumes perfect foreground removal and does not include the modeling of instrumental systematics, which are optimistic assumptions for future observations. To generate more realistic data, the primary beam effects, foreground residuals, and other instrumental systematics can be applied to the simulations before calculating the 2DPS. However, for this work, we keep these simplifying assumptions as a means for proof of concept of our pipeline.

\section{Emulators}\label{emulators}
\subsection{Emulating the 2D power spectrum} \label{sec:emu_2dps}

Emulators provide an alternative method to simulating computationally expensive simulation suites. The emulator itself requires a bank of simulations, but after training can be used for quick, on-the-fly simulation evaluations in an MCMC or SBI framework \citep{Breitman_2023}. Emulators also allow us to improve sampling density across prior space. The goal in building an emulator is to determine whether the emulator, and resulting emulator generated dataset, is accurate enough to be a suitable alternative to the original simulations, and if increasing sampling density of the prior with emulation could improve constraints.  In this case, we consider the simulations as the ground truth, and so assess the SBI pipeline with the emulator relative to these 'true' simulations.

Previously, the 1DPS has been the focus of several works \citep{Schmit_2017, ShimabukuroSemelin, Kern_2017, ghara2020constraining}. We build a 2DPS emulator and apply it in an SBI framework for inference of astrophysical parameters and the ionization history.

\subsubsection{Training of the emulator} \label{sec:eval_metrics}

As mentioned, our emulator was trained on a set of 27,300 21cmFAST simulations, using the three astrophysical parameters as input and predicting the corresponding 2D power spectrum. There are two factors that are taken into account when choosing how to train and evaluate the emulator: the noise from the telescope (in this case we model the expected SKA noise as described in Section \ref{sec:noise}) and the sample variance. To focus the emulator training on regions with a high signal-to-noise ratio (SNR), we divide the 21 cm power spectra by a noise weighting:

\begin{equation}\label{SNR_equation}
    \mathrm{SNR} = \frac{P_{\mathrm{21}}(\textbf{k}, z)}{\sigma_{\mathrm{Noise}}(\textbf{k}, z)},
\end{equation}

where P$_{\mathrm{21}}(\textbf{k}, z$) is the power spectrum of the signal itself, and $\sigma_{\mathrm{Noise}}(\textbf{k}, z) $ is the standard deviation of the noise, as given by Equation \ref{noise_dist_equation}. 

Neural networks train best when given data that has been scaled to reduce dynamical range, so we preprocess our data accordingly \citep{Breitman_2023}. We first divide the simulated power spectra by $\sigma_{\mathrm{Noise}}(\textbf{k}, z) $ to calculate the SNR, we then floor the data at a value of SNR = 0.1, so that the emulator does not focus regions of low signal. Then, after taking the $\mathrm{log}_{10}$ of the SNR data, we scaled the data linearly to have values between 0 and 1. The input astrophysical parameters were also scaled linearly to be between 0 and 1. The data is then flattened into a 1D array, similar to the steps taken in \citep{greig2024inferring}, who also showed that flattening their 2DPS data was sufficient for training their model.

The smallest k-modes, corresponding to the chosen box size of 250 Mpc, are under sampled, leading to a high sample variance in the region of low $k_{\perp}$ and $k_{\parallel}$. Hence, it was not viable to produce a model that predicted well in that region, causing the error in the emulator to exceed the standard deviation of the noise. 

\begin{figure}
	\includegraphics[width=\columnwidth]{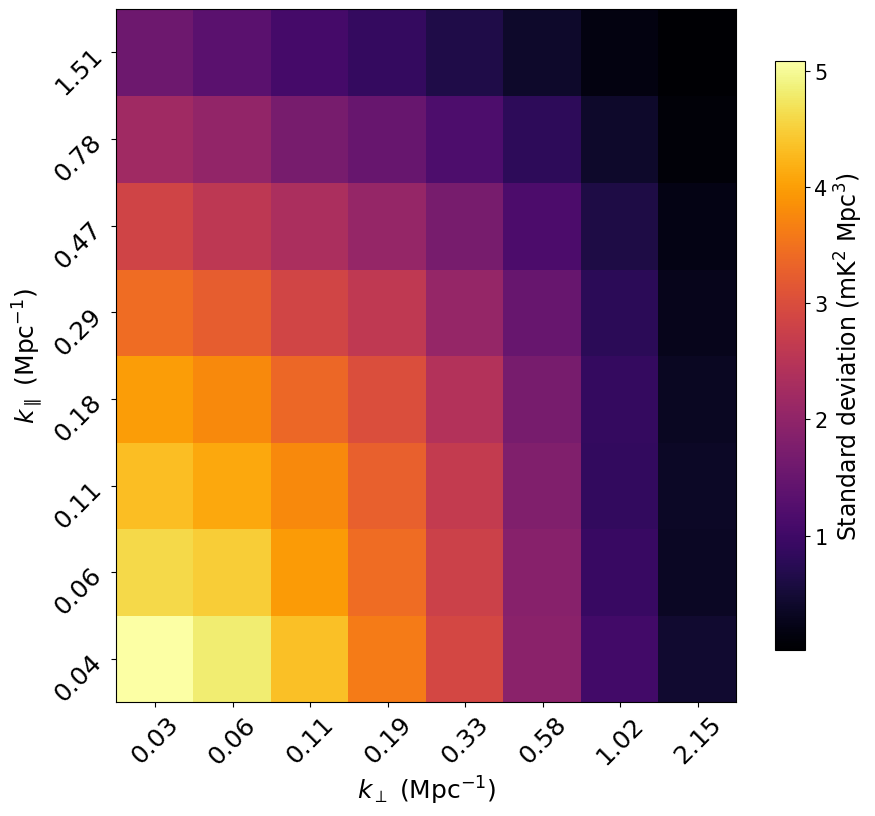}
    \caption{Visualization of the sample variance. Standard deviation of each bin for a 100 simulations of differing initial conditions at z = 7.8 for a fixed parameter sample of: [$\zeta$: 17.99, $\mathrm{log}_{\mathrm{10}}T_{\mathrm{vir}}$(K): 4.82, $R_{\mathrm{mfp}}$(Mpc): 25.96]. This plot is represented in log space.} 
    \label{fig:cosmic_var_emu} 
\end{figure}

To alleviate this problem, we remove some of the pixels in the low $k_{\perp}$ and $k_{\parallel}$ region to gain accuracy whilst minimizing information loss. Fig.  \ref{fig:cosmic_var_emu} reflects the standard deviation per pixel of 100 simulations with both a fixed parameter sample and redshift of z = 7.8. The plot shows the higher standard deviation at the lower k-bins, with a drop off after two bins in the horizontal and vertical direction. There is also a large difference between the bottom left-hand corner pixel and the next diagonal pixel. This trend is seen at all redshifts. For this reason, we found the optimal pixels to remove were the three pixels in the bottom left-hand corner. An example for the parameter sample [$\zeta$: 17.99, $\mathrm{log}_{\mathrm{10}}T_{\mathrm{vir}}$(K): 4.82, $R_{\mathrm{mfp}}$(Mpc): 25.96] for three out of the eight total redshifts (z = 11.2, 7.8 and 5.88) is illustrated in the middle row of Fig.  \ref{fig:square_emu_pred_truth}. Applying this same masking to each simulated power spectrum, the model is then trained on this reduced dataset. 

\begin{figure}
	\includegraphics[width=\columnwidth]{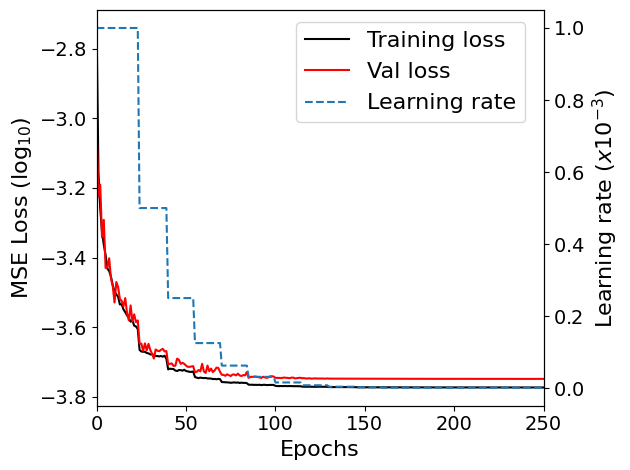}
    \caption{Training loss (black solid line) and validation loss (red solid line) over the training epochs for the emulator of the 2DPS. The learning rate changes are also shown as the dashed blue line. The plot has been clipped to 250 epochs, as training reached convergence.} 
    \label{fig:train-loss-emu} 
\end{figure}

The architecture chosen for the emulator model is shown in Table \ref{tab:example_2dps_emu}. The model consists of 6 fully-connected layers, where the nodes increase to a maximum of 1024 and then decrease down to the size of the output, 8x61. The activation function was chosen as the sigmoid linear unit (SILU) and the model was trained for 500 epochs to ensure convergence. The mean-squared-error loss was chosen as the training loss function, and the learning rate was varied starting from $10^{-3}$ down to $\sim$ $10^{-6}$. The plot of these quantities can be seen in Fig.  \ref{fig:train-loss-emu}, where the validation loss is also shown.

\begin{table}
	\centering
	\caption{Architecture and hyperparameters of the 2DPS emulator, a seven layer fully connected dense neural network}.
	\label{tab:example_2dps_emu}
	\begin{tabular}{lr} 
		\hline
		Layer Type & Activation Function\\
		\hline
        Input, 3 & \\
		Dense, 64 neurons & SILU \\
		Dense, 128 neurons & SILU \\
        Dense, 512 neurons & SILU \\
        Dense, 1024 neurons & SILU \\
        Dense, 1024 neurons & SILU \\
        Dense, 350 neurons & SILU \\
        Output, 488 neurons & Linear\\ 
        %\CN{Output, 488} & \\
		%\hline
        \end{tabular}
        \begin{tabular}{lcr} 
        \hline 
        Optimizer & Loss function & Batch size\\
                \hline
        Adam & MSE Loss & 32
        \end{tabular}

\end{table}

\subsubsection{2D PS emulator performance}\label{emu performance}

To quantify the overall performance of the emulator, we investigated several evaluation metrics. The first was the accuracy of the model, or how well the emulator predicted relative to the true signal. This metric is expressed as follows:  

\begin{equation}
    \mathrm{Accuracy} = \frac{P_{\mathrm{pred}}(\textbf{k}) - P_{\mathrm{truth}}(\textbf{k})}{P_{\mathrm{truth}}(\textbf{k})},
\end{equation}

where $P_{\mathrm{pred}}(\textbf{k})$ is the prediction from the emulator and $P_{\mathrm{truth}}(\textbf{k})$ is the true signal. 

The next two metrics, the error relative to the noise and sample variance contributions, follow from the analysis in \cite{mcquinn2006cosmological}. The error in the emulator relative to the standard deviation of the noise is represented as:
 
\begin{equation}\label{noise_weighted_error}
    \text{Error}_{\text{Noise}} = \frac{P_{\mathrm{pred}}(\textbf{k}) - P_{\mathrm{truth}}(\textbf{k})}{\sigma_{\mathrm{Noise}}(\textbf{k}, z)},
\end{equation}
where the denominator is the standard deviation of the noise, as described in Equation \ref{SNR_equation}.

Finally, to take into consideration the sample variance (SV) the error in the model was calculated with respect to the noise plus the SV, as shown in Equation \ref{SV-noise-err} below.

\begin{equation}\label{SV-noise-err}
    \text{Error}_{\text{SV, Noise}} = \frac{P_{\mathrm{pred}}(\textbf{k}) - P_{\mathrm{truth}}(\textbf{k})}{\sigma_{\mathrm{Noise}}(\textbf{k}, z) + \sqrt{\frac{2}{N_{\mathrm{modes}}}}P_{\mathrm{truth}}(\textbf{k})}
\end{equation}

The denominator contains both the noise variance contribution, $\sigma_{\mathrm{Noise}}(\textbf{k}, z)$, and the sample variance contribution, $\sqrt{\frac{2}{N_{\mathrm{modes}}}}P_{\mathrm{truth}}(\textbf{k})$ \citep{mcquinn2006cosmological}.

\begin{figure}
	\includegraphics[width=\columnwidth]{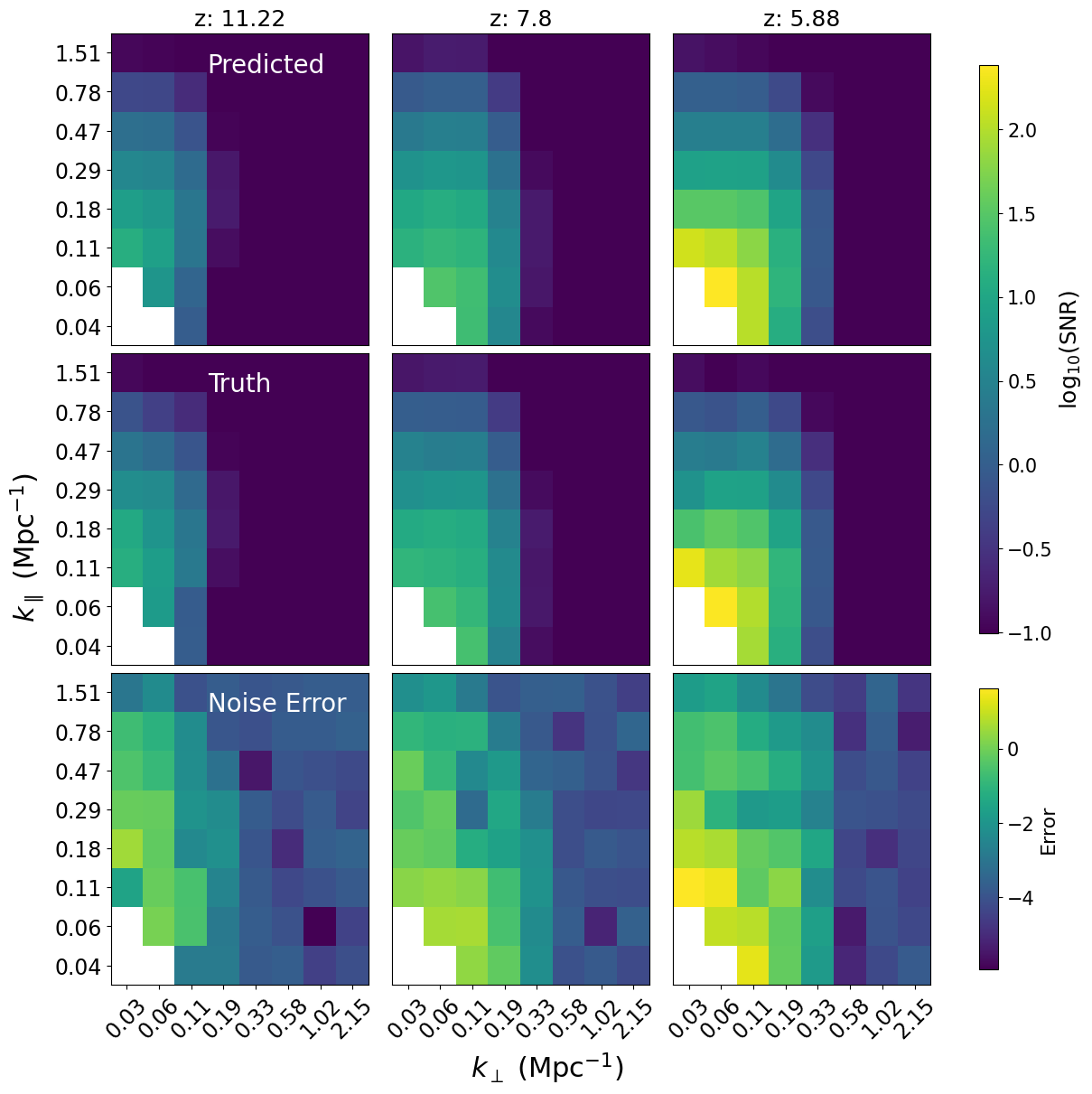}
    \caption{Example prediction from the 2D power spectrum emulator at the fiducial point  [$\zeta$: 17.99, $\mathrm{log}_{\mathrm{10}}T_{\mathrm{vir}}$(K): 4.82, $R_{\mathrm{mfp}}$(Mpc): 25.96]. The top row shows the emulator predicted 2DPS, the middle row shows the truth 2DPS, and the bottom row shows the error relative to the spread of the noise. The error is calculated as shown in Equation \ref{noise_weighted_error}. The emulator was trained on data with the three bottom left hand corner pixels removed, which are represented above with the white masking. For conciseness, only 3 out of the 8 redshifts are shown. The power spectra are given in the signal-to-noise ratio with k$_\perp$ bins on the bottom x axis and k$_\parallel$ bins on the y axis.}
    \label{fig:square_emu_pred_truth} 
\end{figure}

To check that the training dataset for our emulator contained sufficient samples to reach convergence, we halved the training set from $27,300$ to $13,650$ and found that the emulator accuracy error did not decrease, indicating we had trained our emulator on enough samples.
An example prediction from the 2D power spectrum emulator can be seen in Fig.  \ref{fig:square_emu_pred_truth}. The point in parameter space associated with this plot is [$\zeta$: 17.99, $\mathrm{log}_{\mathrm{10}}T_{\mathrm{vir}}$(K): 4.82, $R_{\mathrm{mfp}}$(Mpc): 25.96]. The top row shows the emulated 2D power spectra at three out of the eight total redshifts for ease of viewing (z = 11.22, 7.8, 5.88). The middle row shows the truth power spectra, and the bottom row shows the noise weighted error between the two, as calculated with Equation \ref{noise_weighted_error}. Visually, the predicted spectra appear with similar structure and power range as that of the truth signal. The bottom row shows pixels in the lowest redshift plot with errors that are orders of magnitude above the standard deviation of the noise. We attribute this to the challenge of capturing the sample variance at those $k$ bin scales. 

\begin{figure}
	\includegraphics[width=\columnwidth]{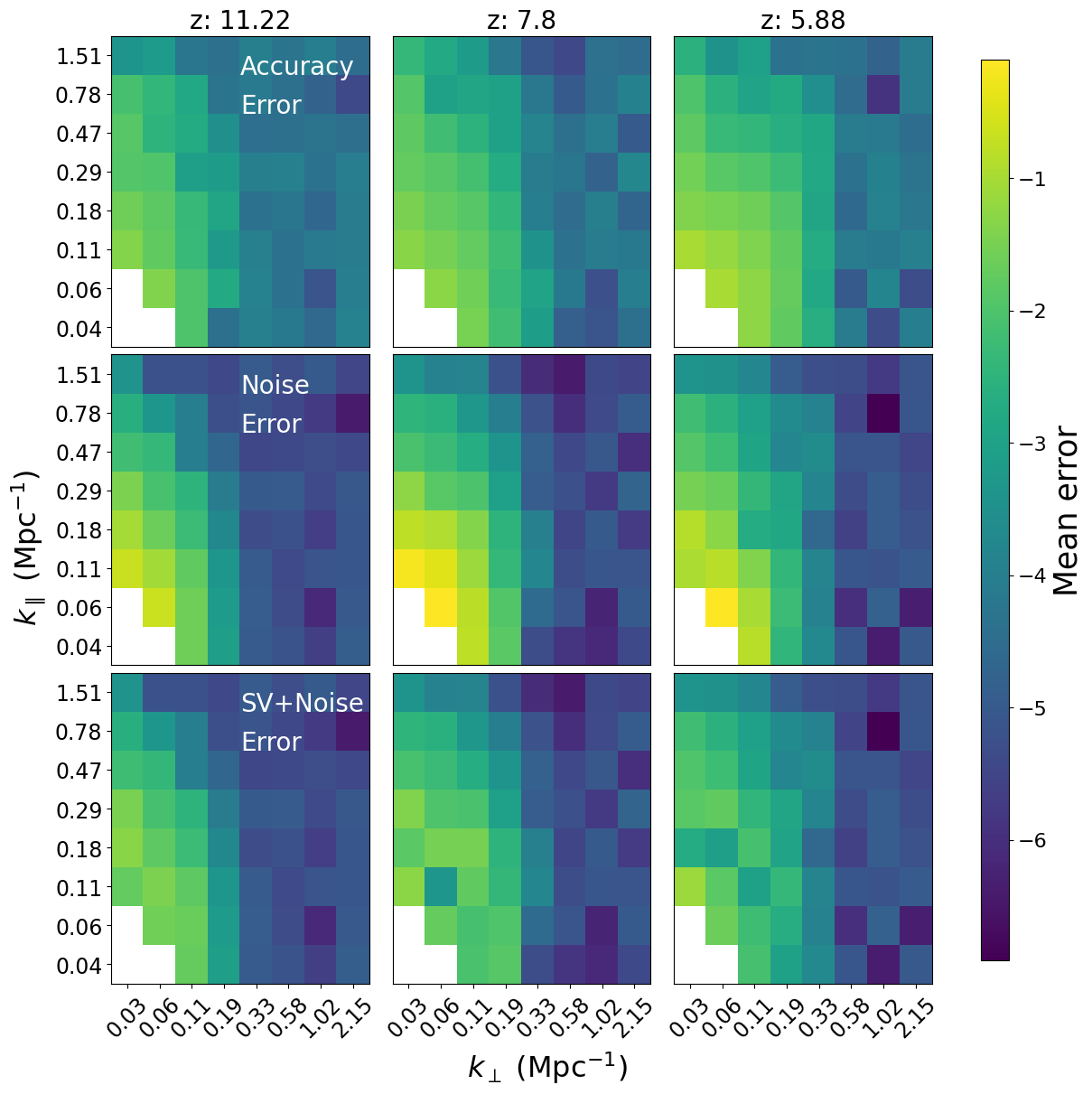}
    \caption{Plot of the three measures of mean error. The top row shows the accuracy, the middle row shows the noise weighted error, and the bottom row shows the sample variance and noise weighted error. For conciseness and consistency, only three out of the eight redshifts are shown, matching those in Fig.  \ref{fig:square_emu_pred_truth}. Each of the errors are calculated as shown in Section \ref{sec:eval_metrics} and are each represented as log$_{10}$ of the fractional error.} 
    \label{fig:grid_bias_err_emu} 
\end{figure}

We split the error analysis into mean error and standard deviation for each of the three metrics described in the previous section. Fig.  \ref{fig:grid_bias_err_emu} shows the mean error with respect to the accuracy, noise weighting, and noise and SV weighting. The mean error is calculated by taking the absolute value of the mean of the error in each pixel, and shows in each redshift bin how far on average the values vary from the mean. The top and bottom rows in Fig.  \ref{fig:grid_bias_err_emu} show that the mean error in accuracy and SV + noise error fall well beneath the 10\% level, approaching the 1\% error across most of the image. The mean noise error in the middle row shows that the pixels around those that have been masked are near the 100\% level, indicating that the model error is larger than the noise standard deviation.
Including the SV weighting component, in addition to the noise, surpasses the model error, indicating the model performs well relative to the SV. Evidently, structure still remains in the mean error with all three evaluation metrics, with highest error power in the lower left-hand corner, where the highest 21cm signal power also exists. The residual structure reflects the pattern seen in the sample variance in Fig.  \ref{fig:cosmic_var_emu}. We can only expect the emulator to be able to predict the mean, so deviations from the truth signal generated with a random seed are inevitable.

Due to the deterministic nature of the emulator, it is not possible to predict the distribution of power in the low $k$ bins. To investigate further, we increased the number of seeds per parameter combination to six total and re-trained the model on all the seeds to check if the emulator would be able to better capture the SV in the low $k$ bins by predicting the mean. The resulting model did not perform better than the model trained on one seed. We also trained several emulators for each seed set of data and averaged over the results, this also did not improve the results. In addition, since the 2DPS can be analysed as an image, we trained a convolutional neural network (CNN) on the image format of the data to assess whether the CNN improved feature extraction. However, the CNN also did not improve performance. Hence, we decided to proceed with the simplest model.

\begin{figure}
	\includegraphics[width=\columnwidth]{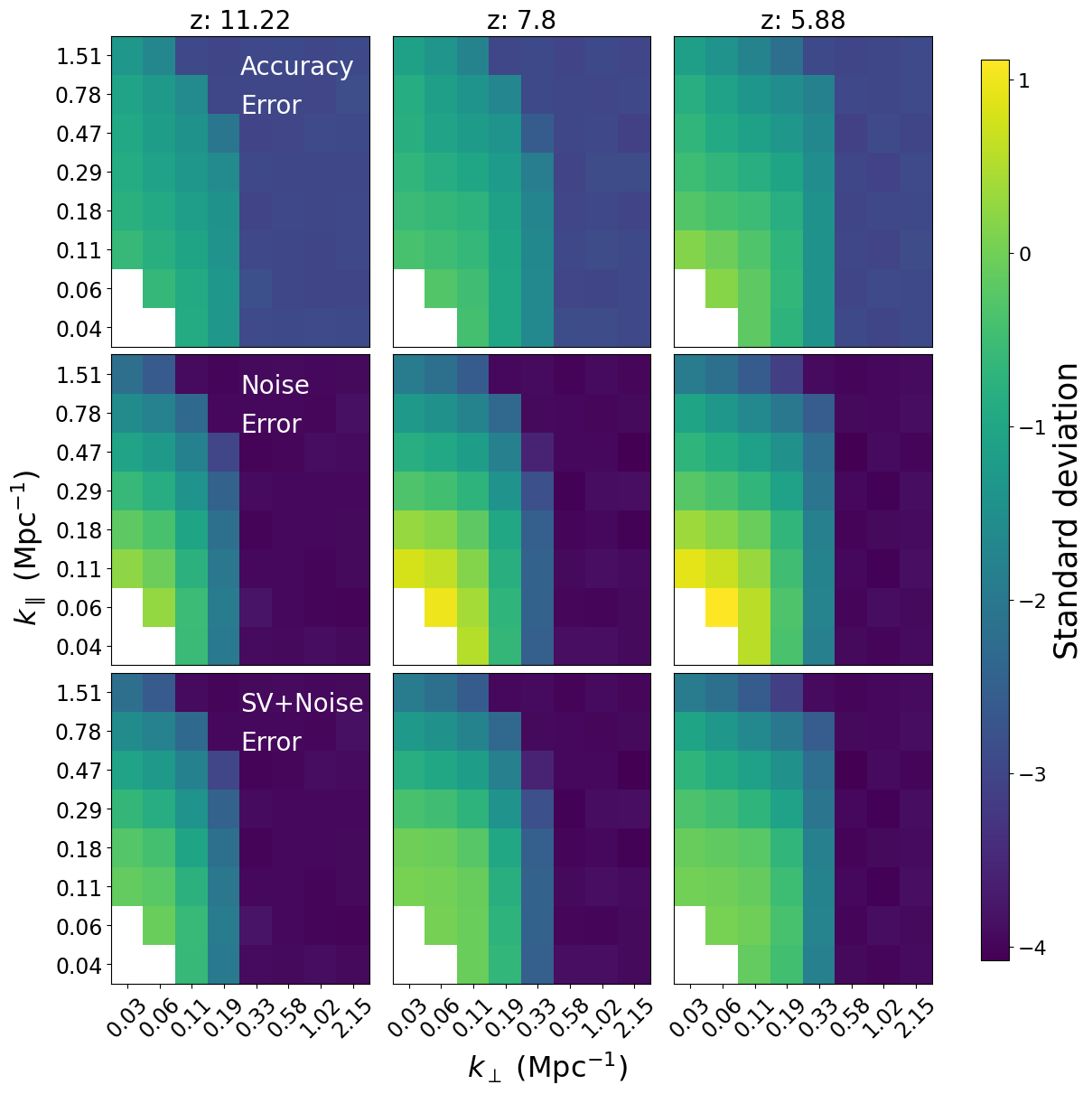}
    \caption{Plot of the standard deviation of the three error metrics. The top, middle, and bottom row show the standard deviation of the accuracy, noise, and noise and SV error, respectively. Each error is calculated as shown in Section \ref{sec:eval_metrics} and are each represented as log$_{10}$ of the fractional error. All of these errors are dimensionless quantities.} 
    \label{fig:grid_mag_err_emu} 
\end{figure}

In Fig.  \ref{fig:grid_mag_err_emu}, we show the standard deviation of the error in each redshift bin. This was calculated by taking the standard deviation of the absolute value of the error as prescribed by the three evaluation metrics. We see a similar pattern to the mean error plot, where the top and bottom plots hover closer to the 10\% error level showing lower errors than the middle row. The standard deviation of the noise error jumps above 100\% error in the pixels closest to the masked corner, again showing that the noise spread on its own does not cover the error in the emulator. Once the SV is taken into account, the standard deviation decreases to 10\% or under. 

\cite{Kern_2017} emulate the 1DPS using Principal Component Analysis (PCA) and Gaussian Process Regression (GPR). They are able to emulate at $\leq$ 5\% precision across most of their power spectra predictions. The emulator built by \cite{Breitman_2023}, 21cmEMU, emulates the 1DPS using a dense neural network and reaches sub-percent fractional error in most of the prediction space, increasing to a maximum of $\sim$ percent levels. \cite{meriot2025comparison} train a dense neural network for their 1DPS emulator and are able to get $\sim 3 \%$ accuracy. So while we reach under 10\% error for the SV+noise error, the 1DPS emulator models are performing with more accuracy. This is likely due to the 2DPS experiencing a higher sensitivity to sample variance since there are fewer independent Fourier modes per bin. In contrast, the 1DPS spherically averages over the Fourier modes, decreasing the variance in each $k$ bin.

\subsection{Emulating the Ionization History}

As described in Section \ref{SBI}, we use neural posterior estimation in order to infer the astrophysical parameters of our model. To obtain the ionization history, we draw parameter combinations from the inferred posterior distribution at a given observation. With 21cmFAST, we can forward model to obtain the ionization history at each of these parameter combinations. However, as mentioned in Section \ref{emulators}, running large numbers of 21 cm simulations is computationally expensive. Hence, we are limited in the number of parameter draws at which we can simulate the corresponding ionization history. This would lead to under-sampling of the posterior distribution for the ionization history. 

\begin{figure}

	\includegraphics[width=\columnwidth]{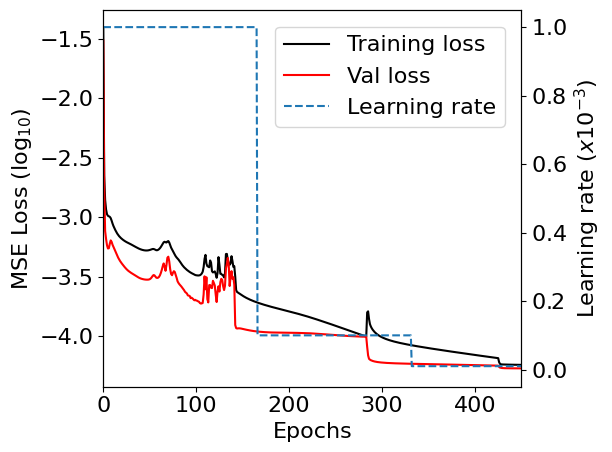}
    \caption{Training loss (black solid line) and validation loss (red solid line) over the training epochs for the emulator of the ionization history. The learning rate changes are also shown as the dashed blue line. The plot has been clipped to 450 epochs, as training reached convergence.}
    \label{fig:train-loss-emu-ion} 
\end{figure}

We, therefore, train an emulator of the ionization history from our model parameters to the neutral fraction in 41 redshift bins. We use an architecture with eight fully connected layers and a Gaussian error linear unit (GELU) activation function. The complete architecture and hyperparameters are shown in Table \ref{tab:emu_ion_arch}. The model was trained with a variable learning rate, starting at $10^{-3}$ and dropping to $10^{-6}$  which can be seen in Fig. \ref{fig:train-loss-emu-ion}, along with the training and validation loss.

\begin{table}
	\centering
	\caption{Architecture and hyperparameters of the ionization history emulator, a nine layer fully connected dense network.}	\label{tab:emu_ion_arch}
	\begin{tabular}{lr} 
		\hline
		Layer Type & Activation Function\\
		\hline
            Input, 3 \\
		Dense, 64 neurons & GELU \\
		Dense, 112 neurons & GELU \\
            Dense, 256 neurons & GELU \\
            Dense, 512 neurons & GELU \\
            Dense, 1024 neurons & GELU \\
            Dense, 1024 neurons& GELU \\
            Dense, 512 neurons & GELU \\
            Dense, 256 neurons & GELU \\
            Output, 41 neurons & Linear \\
		\hline
        \end{tabular}
        \begin{tabular}{lcr} 
        \hline 
        Optimizer & Loss function & Batch size\\
                \hline
        Adam & MSE Loss & 1,000
        \end{tabular}

\end{table}

\begin{figure}

	\includegraphics[width=\columnwidth]{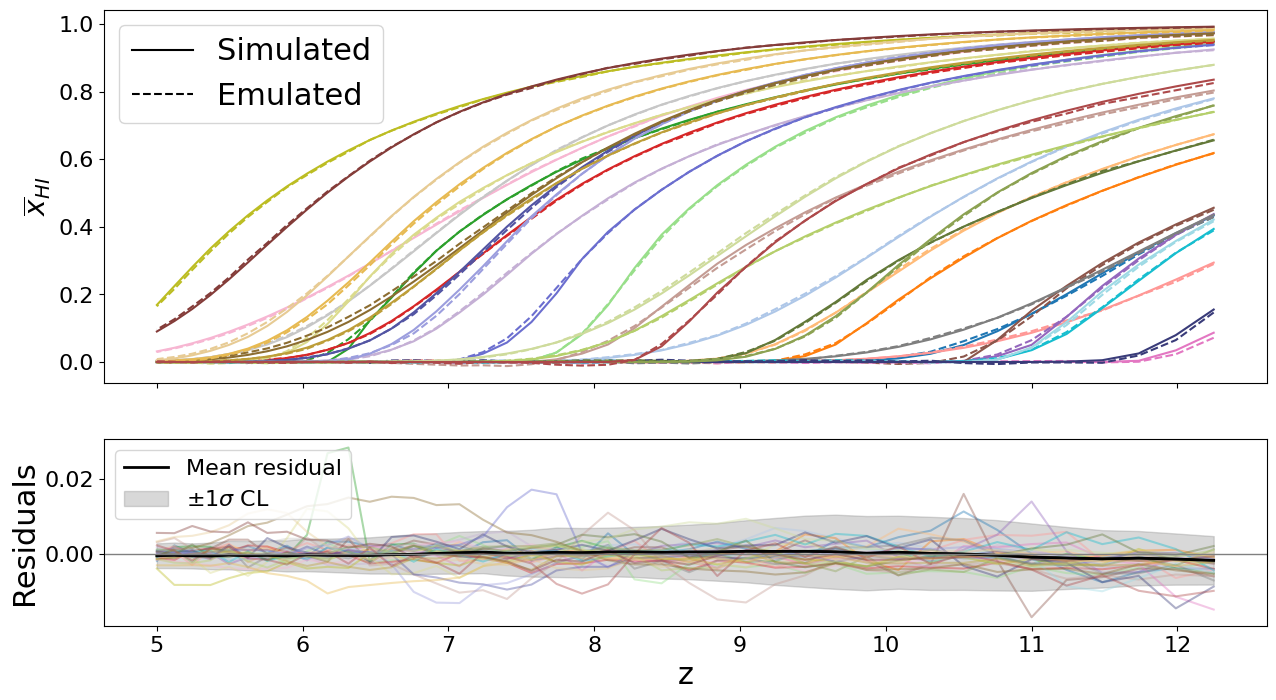}
    \caption{Simulated and emulated ionization histories. The solid lines represent the simulated ionization histories and the corresponding colored dashed lines represent the emulated ionization history for the same parameter combination. The bottom panel shows the residuals for the corresponding prediction. The shaded region indicates 1$\sigma$ confidence level.}
    \label{fig:emulator_ion_hist} 
\end{figure}

Fig.  \ref{fig:emulator_ion_hist} shows a randomly selected collection of simulated and emulated ionization histories from the test set. The test set is completely unseen in the training process. Visually, you can see good agreement between the simulator and the emulator. We calculate the mean and standard deviation of the residual error at each redshift. This is shown by the shaded region in the bottom panel. We chose to show the residuals, rather than fractional error, to allow us to express error in parts of parameter space where the neutral fraction is 0.

\section{Simulation Based Inference}\label{SBI}

Simulation based inference \citep{cranmer2020frontier} (also known as implicit likelihood inference (ILI) and likelihood-free inference (LFI)) is a technique  for learning the statistical relationship between observed data, \textit{d}, and model parameters, \textit{$\theta$}. SBI trains a neural network to best fit the parameters of some surrogate distribution to the likelihood or posterior distribution across parameter space. 

Using a model which captures all the physics of your observable; such as noise, instrumental effects, and the signal itself, you are able to generate simulations of the observation at many parameter combinations. Drawing a parameter combination from your prior and forward modeling to obtain mock data is equivalent to sampling from the joint distribution of the parameters and data. These samples can be used as a training bank to fit the likelihood or posterior over the whole of prior space. In this way, you circumvent the need for an implicit form of the likelihood. This is beneficial as we often do not have an analytic form of the exact likelihood. 
In parameter inference from the 21 cm signal, the likelihood is often assumed to be Gaussian; see \cite{greig201521cmmc, ghara2020constraining, maity2023efficient}.  As discussed in \cite{prelogovic2023exploring}, despite primordial density fluctuations being Gaussian, non-linear evolution, such as gravitational collapse, leads to non-Gaussianities in the hydrogen density field. Furthermore, the temperature and ionization state of the IGM is driven by radiation fields of the first galaxies, leading to non-trivial mode coupling \citep{prelogovic2023exploring}. Due to this unknown dependence, it is intractable to find an exact functional form of the 21 cm likelihood, leading to the common Gaussian approximation. \cite{prelogovic2023exploring} find that a Gaussian likelihood results in an overconfident biased posterior. Conversely, \cite{meriot2025comparison}, for the 1D PS, find that a Gaussian likelihood is sufficient for a 100h SKA noise level, as the uncertainty due to noise outweighs the non-Gaussianities. SBI models the likelihood implicitly, providing an alternative to the Gaussian likelihood approximation; see \cite{zhao2022simulation, saxena2024simulation, greig2024inferring, meriot2025comparison}.

Given a large enough bank of data-parameter pairs $\boldsymbol{(d, \theta)}$ sampled from the joint distribution, or a fast forward model of our observables, we can use a neural density estimator to find the best fit to either the likelihood $\boldsymbol{P(x\mid\theta)}$ (Neural Likelihood Estimation  (NLE)) or the posterior $\boldsymbol{P(\theta \mid x)}$ (Neural Posterior Estimation (NPE)) \citep{lueckmann2017flexible, papamakarios2019neural}. Neural networks are trained to emulate probability distributions. Common choices of neural density estimator include mixture density models (MDNs) \citep{bishop1994mixture} or masked autoregressive flows (MAFs) \citep{papamakarios2021normalizing}. The neural network is trained to specify the parameters of the MDN, such as means and variances of Gaussians. In the case of flows, the network is trained to learn a set of invertible transformations of a base Gaussian into a more complicated target distribution. 

\subsection{Neural Posterior Estimation} 
For this work, we use a technique called neural posterior estimation \citep{papamakarios2019neural}. Our neural network architecture is learning an approximate posterior $\boldsymbol{\hat{P}(\theta\mid x)}$ from a set of data-parameter pairs in our training set $\boldsymbol{(x, \theta)}$. Here, our loss function is the negative log-posterior of our learnt neural density estimator $\boldsymbol{\hat{P}(\theta \mid x)}$ \citep{papamakarios2016fast}:

\begin{equation}
\begin{aligned}
    \mathcal{L}_{\text{NPE}} &= - \mathbb{E}_{\mathcal{D}_{\text{train}}} \left[ \log \hat{P}(\theta \mid x) \right] \\
    &= - \mathbb{E}_{\mathcal{D}_{\text{train}}} \log \left[  {p(\theta)}q_{w}(\theta \mid x)  \right],
\end{aligned}
\end{equation}

where $\boldsymbol{\hat{P}(\theta \mid x)}$, is divided into $q_{w}(\theta \mid x)$, the output of the neural network and the prior distribution ${p(\theta)}$. 

The advantage of NPE is that inference can be repeated for many test observations rapidly. When the likelihood is modeled, MCMC chains must be run in order to obtain samples from the posterior. Obtaining fast inference at many parameter combinations also allows fast evaluation of posterior performance tests over all of the prior volume. 

\subsection{Inference of Astrophysical Parameters} \label{sec:inference}

We used neural posterior estimation to infer astrophysical parameters from noisy 2DPS mock observations. Inference was performed using the Learning the Universe - Implicit Likelihood Inference package, \citep{ho2024ltu}, with the SBI backend \citep{tejero-cantero2020sbi, BoeltsDeistler_sbi_2025}. The inference pipeline consisted of an ensemble of a neural spline flow \citep{durkan2019neural} and a masked autoregressive flow \citep{papamakarios2021normalizing}. The final posterior density was obtained by sampling from the first two architectures seen in Table \ref{tab:SBI_arch}. Each of the architectures were independently trained, with the same training set. Samples were then drawn from each architecture and averaged over to obtain the final posterior estimate. This averaging is weighted by the value of each model's validation loss, favoring higher performing models. This is to mitigate any bias that a particular architecture may introduce.

In order to select the best architecture for the neural density estimation, 5 architectures were first independently trained shown in Table \ref{tab:SBI_arch}. These were comprised of a MAF \citep{papamakarios2021normalizing}, a neural spline flow \citep{papamakarios2021normalizing}, a 10 component MDN  \citep{bishop1994mixture}, a 3 component MDN \citep{bishop1994mixture}, and a Masked Autoencoder for Distribution Estimation (MADE) \citep{germain2015mademaskedautoencoderdistribution}. We assessed the performance of each architecture using SBC. We compared the SBC of the individual networks with an ensemble network weighted by the value of the final validation loss. We iteratively included the next highest performing architecture, until adding more architectures stopped improving the overall performance of the ensemble. This resulted in an ensemble comprised of the first two architectures in Table \ref{tab:SBI_arch}. For further discussion of the ensemble, see Appendix \ref{architectures}.

Two inference pipelines were compared. The first pipeline consisted of a training set of data-parameter pairs generated using 21cmFAST. The second pipeline was a training set consisting of samples generated with the emulator of the 2DPS described in Section \ref{sec:emu_2dps}.

\begin{table}
	\centering
	\caption{Architecture and hyperparameters of each of the architectures investigated. Only the top two were included in the final neural posterior ensemble.}\label{tab:SBI_arch}
	\begin{tabular}{lcr} 
		\hline
		Model & Number of transforms/components & Hidden Features\\
		\hline
		MAF & 12 & 50\\
		NSF & 5 & 50 \\
        MDN & 10 & 50 \\
        MDN & 3 & 50 \\
        MADE & 5 & 50 \\
		\hline
        \end{tabular}
        \begin{tabular}{lcr} 
        \hline 
        Learning rate & Loss function & Batch size\\
                \hline
        $ 1e^{-4}$ & neg log prob & 1,000
        \end{tabular}

\end{table}

Our first inference method was trained on the $8 \times 8 \times 8 $ dimensional power spectra generated using 21cmFAST, and their corresponding parameters. Before training, the data was first preprocessed. For each simulation in the training set, 50 noise realizations were generated to obtain 50 mock observations per parameter combination. This was to allow the SBI method to learn the uncertainty due to the spread of the noise. The mean noise power spectra was then subtracted to form noisy observations. Each noisy observation was then divided by the standard deviation of the noise per pixel, given by Equation \ref{noise_dist_equation}.

The data was then preprocessed in the same way as described in Section \ref{sec:emu_2dps}. This choice was made in order to force the neural network to focus on regions with the highest signal-to-noise. The SBI framework consists of an embedding network, \citep{BoeltsDeistler_sbi_2025} trained jointly with a neural posterior estimator with the hyperparameters in Table \ref{tab:SBI_arch}. Since they are trained jointly, the SBI framework still has access to the full information contained within the summary. The embedding network acts to reduce the dimensionality which the neural density estimator has to fit. The architecture of this embedding net can be found in Table \ref{tab:densenet}. A visualization of how consistently the architectures perform with respect to one another can be seen in Appendix \ref{architectures}. We use multiple architectures for the neural surrogate posterior density to remove dependency on the model choice. There is some discrepancy between models, but we mitigate this bias by averaging over the architectures.

\begin{table}
	\centering
	\caption{Architecture of the embedding net.}
	\label{tab:densenet}
	\begin{tabular}{lr} 
		\hline
		Layer Type & Activation Function\\
		\hline
		Dense, 512 neurons & Tanh \\
		Dense, 1024 neurons & Tanh \\
            Dense, 1024 neurons & Tanh \\
            Dense, 512 neurons & Tanh \\
            Dense, 6 neurons & Linear \\
		\hline
        \end{tabular}
\end{table}

The second pipeline consisted of our neural density estimator on data parameter pairs obtained with our emulator. We drew 1M training samples from a uniform prior. Each of these samples has a unique noise realization. Our emulator is a deterministic process, whereas the true 2DPS is stochastic due to sample variance. We want the neural density method to be able to learn this stochasticity. We, therefore, approximate the spread due to sample variance as \citep{mcquinn2006cosmological},

\begin{equation} 
    \delta P_{\mathrm{SV}}(\textbf{k},z) = \mathcal{N}\left(\mu=0, \sigma^2 = \frac{2 \hat{P}_{\mathrm{21}}^{2}(\textbf{k},z)}{N_{\mathrm{modes}}(\textbf{k},z)}\right).
\end{equation} 

We then draw a sample variance realization in addition to the noise realization. This approximation assumes that the distribution of the sample variance is centered on the emulator prediction. In order to maximize sampling density of the parameters, each point in parameter space was only simulated at one random seed, meaning this may not be an accurate assumption. Furthermore, modeling sample variance in this way is equivalent to a Gaussian likelihood approximation, which may not be a valid assumption as discussed in Section \ref{sec: intro}. The noise and sample variance realisation are added to the output of the emulator in order to generate samples for the SBI pipeline. Training on emulated samples without sample variance realisations added to the neural network output resulted in narrow overconfident constraints.  The neural density estimation architecture is kept consistent between the two pipelines. The test set for both pipelines was generated as in Section \ref{data} using 21cmFAST, not the emulator, and is unseen by the training process. We are working under the assumption that the output from 21cmFAST is our truth. The test set is comprised of simulations from 21cmFAST and not the emulator, since we want to assess the ability of our pipeline to recover the ground truth. This is to determine if the 21cmFAST code can be replaced by a faster emulator. If we evaluate on a test set with emulated samples, but the emulator isn't correctly capturing the simulations, this would inflate the perceived performance of the emulator trained pipeline. We found that training the SBI network with multiple initial condition seeds per parameter combination did not improve performance.

\begin{figure}
	\includegraphics[width=\columnwidth]{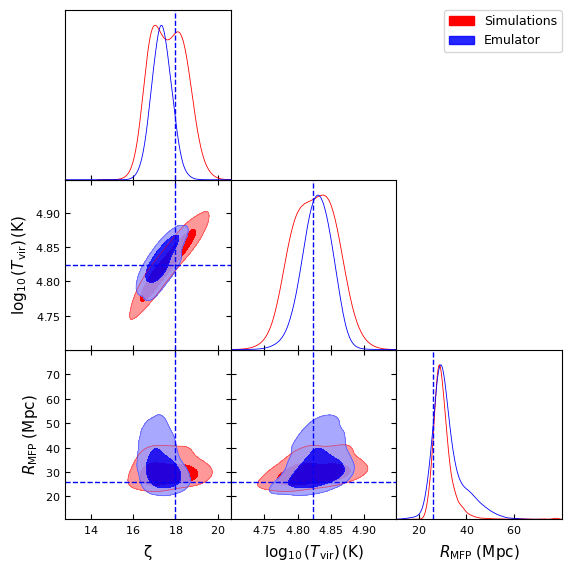}
    \caption{Inference on noisy 21cmFAST 2DPS drawn from the test simulation bank. The red contours show the SBI pipeline trained on simulations and the blue contours show the obtained constraints from the SBI pipeline trained on emulated samples. The posteriors show the one and two sigma contours for 10,000 posterior samples. The ground truth used to simulate the observation is given by the dashed lines.}
    \label{fig:inference_params_emu} 
\end{figure}

\begin{figure}
 	\includegraphics[width=\columnwidth]{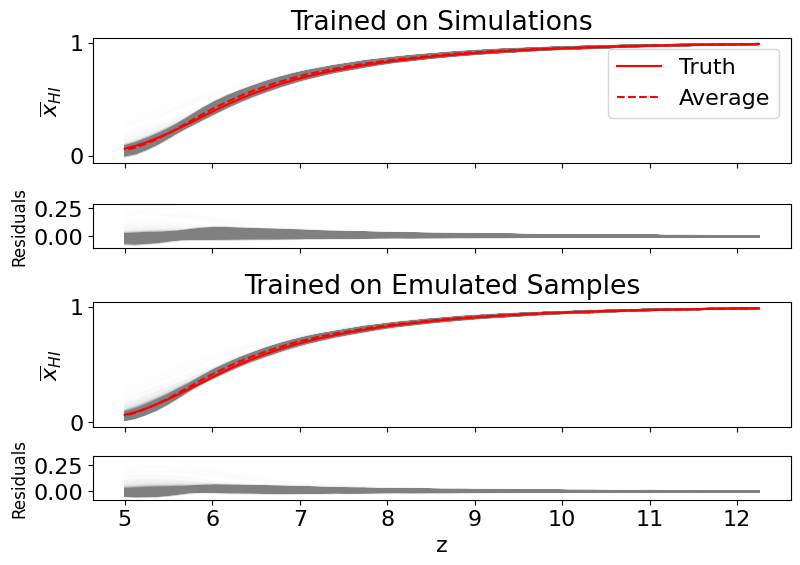}
    \caption{The top panel shows the inferred ionization history for our inference method trained on simulations. The bottom panel shows the inferred ionization history for our inference method trained on emulated samples. The solid orange line shows the target ionization history and the dashed line shows the mean of ionization history samples. The cloud of gray points shows 10,000 samples from the posterior distribution of our ionization history. These ionization histories are for the corresponding parameter combination to Fig.  \ref{fig:inference_params_emu}.} 
    \label{fig:inference_ion_emu} 
\end{figure}

A comparison between constraints obtained with both pipelines is shown in Fig.  \ref{fig:inference_params_emu}. There is a degeneracy between the ionizing efficiency and the virial temperature which is recovered in both inference pipelines. We see good agreement between obtained constraints in both pipelines. We obtain 10,000 posterior samples and emulate the ionization history at each of these parameter combinations. Fig.  \ref{fig:inference_ion_emu} shows the inferred ionization history at each of the parameter combinations for the two methods. The truth ionization history and average of the samples are also shown. Both are able to accurately recover the ionization history, with comparable accuracy. The posterior distribution for the ionization history at eight redshifts can be seen in Fig.  \ref{fig:ion_corner_plots}. The recovered ionization histories for the two pipelines  for the corresponding observation can be seen in Fig.  \ref{fig:inference_ion_emu}. The parameter constraints obtained with the SBI pipeline trained on the emulated samples are comparable to the constraints obtained with the simulations, indicating good agreement between these methods.

We quantify the error in the performance of the ionization history emulator to be less than  $\pm 0.01$ on the neutral fraction in nearly all cases. The size of this systematic uncertainty is smaller than the posterior width, at most redshifts. At the lowest redshifts, where the gas is close to fully ionized, the residual can become comparable to the width of the posterior. In this regime, the statistical error is likely comparable to model errors, which struggle to capture this very ionized regime. For this reason, we conclude the ionization history emulator performs satisfactorily for the purpose of inference.

\subsection{Early vs Late Reionization}
We find much more constraining power for late reionization scenarios when compared to early reionization. Measurements of the CMB have been used to constrain the midpoint of reionization ($z_{re} = 7.68 \pm 0.79) $, using a simple tanh model \citep{aghanim2020planck}. This indicates that the early reionization model is less physically motivated, meaning we recover tight constraints in the parts of parameter space we are most interested in. 

As described in Equation \ref{frequency noise scaling}, the noise scales with the sky temperature which scales with frequency. This introduces a redshift dependence on the noise magnitude. There is much lower SNR at high redshift for this reason. For early reionization, the 2DPS consists of low SNR in the highest redshift bins, and no signal in any bins after reionization takes place and the neutral fraction has dropped to zero. 

\begin{figure}
	\includegraphics[width=\columnwidth]{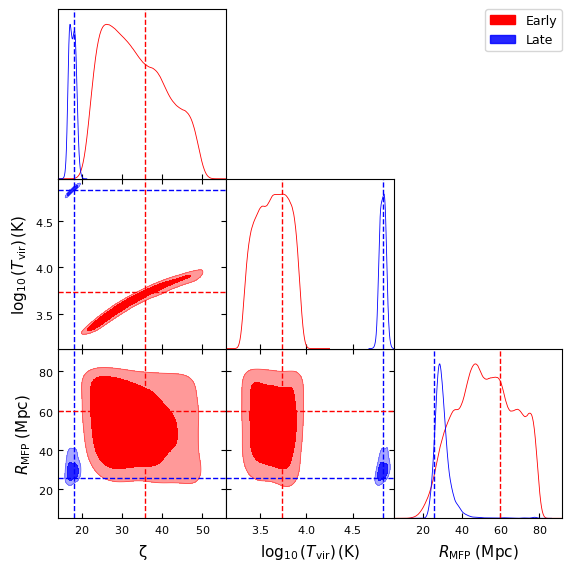}
    \caption{Comparison of inference on noisy 21cmFAST 2DPS for an early vs late reionization case. The red contours show our constraints for an early reionization case, whereas the blue contours are the constraints obtained for a late reionization. The posteriors show the one and two sigma contours for 10,000 posterior samples. The ground truth used to simulate the observations is given by the dashed lines.}
    \label{fig:early_vs_late_inference} 
\end{figure}

\begin{figure}
	\includegraphics[width=\columnwidth]{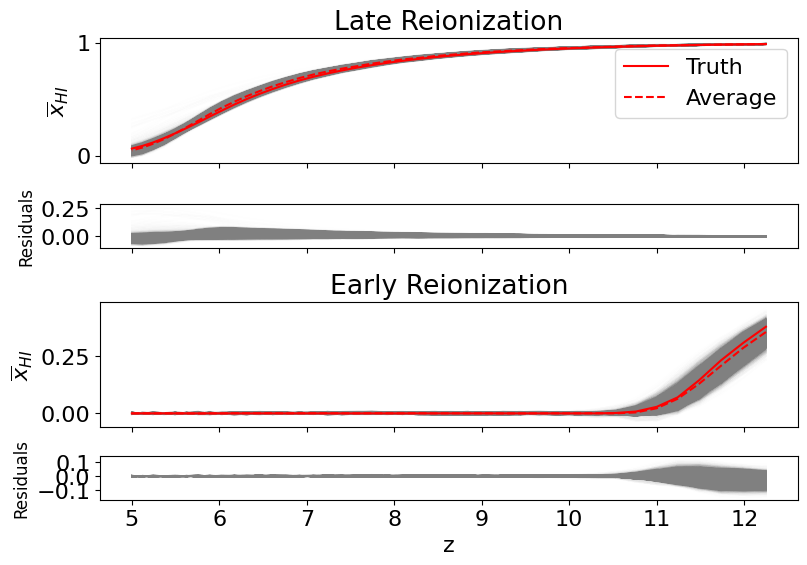}
    \caption{Inferred ionization history for our inference method trained on simulations in the case of late and early reionization. The solid red line shows the target ionization history and the dashed red line shows the mean of ionization history samples. The cloud of gray points shows 10,000 samples from the posterior distribution of our ionization history.} 
    \label{fig:early_vs_late_inferred_ion_hist} 
\end{figure}

These factors combined, therefore, mean that the network has much less information in the cases of early reionization. This results in wider contours. A comparison between constraints for early and late reionization cases can be seen in Fig.  \ref{fig:early_vs_late_inference}.  The inference on an early reionization observation shows much weaker constraints, indicating there is much less information in these signals. The corresponding ionization histories for these parameter combinations can be seen in Fig.  \ref{fig:early_vs_late_inferred_ion_hist}. As can be seen, the agreement between the truth and obtained ionization history is less accurate in the case of early reionization. The constraints on the ionization history are also much wider at higher redshift. This is observed with both training sets, but is plotted only with the simulated training dataset for clarity.

\subsection{Comparison Between Pipelines}
\begin{figure}
	\includegraphics[width=\columnwidth]{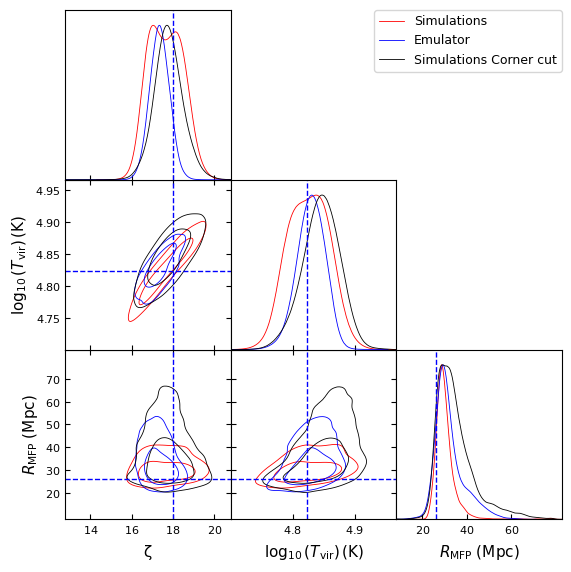}
    \caption{Inference on noisy 21cmFAST 2DPS drawn from the test simulation bank. The red contours show the SBI pipeline trained on the full simulations and the blue contours show the obtained constraints from the SBI pipeline trained on emulated samples. The black contours show the contours obtained from the SBI pipeline with the corner removed. The ground truth used to simulate the observation is given by the dashed lines.}
    \label{fig:no_corner_corner} 
\end{figure}

Inference was performed on a training set of data parameter pairs obtained via 21cmFAST and via the emulator of the 2DPS described in Section \ref{sec:emu_2dps}. An example inference at the same fiducial point in parameter space as in Fig.  \ref{fig:inference_params_emu} can be seen in Fig.  \ref{fig:no_corner_corner} for both of these datasets. We note that the SBI method trained on simulations demonstrates better parameter recovery, as supported by coverage tests described in Section \ref{SBC}. There are two possible causes for the pipeline's performance displaying a preference towards training on simulations. The first, loss of information when removing the corner, and the second  model error introduced by the emulator. In order to determine which cause impacted the results most significantly, we retrained the inference pipeline on simulations with the same pixels cut as in the emulator. Fig.  \ref{fig:no_corner_corner}, includes all three posterior distributions for easy comparison. All of the trained estimators recover the degeneracy between the ionizing efficiency and the virial temperature.

\begin{figure}
        \hspace{0.8cm}
	\includegraphics[width=0.7\columnwidth]{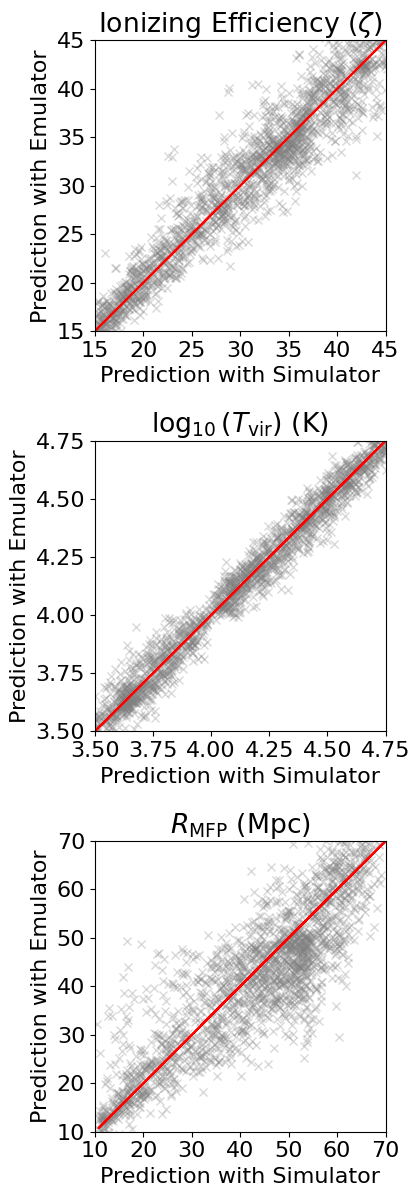}
    \caption{A measure of the agreement between the predictions between the SBI trained on simulations and the SBI trained on emulated samples. The crosses represent the mean value of 1,000 samples from a posterior evaluation of each pipeline for each of the 1,986 simulations in the test set.}
    \label{fig:emu_vs_sim} 
\end{figure}

In order to quantify the agreement between the two SBI pipelines, we calculated the mean of 1,000 posterior samples at each of the test inferences in both cases. The test set was generated as described in Section \ref{data}. We remove samples from our test set which fall at the edge of the prior. This is to ensure inferences at these points are not introducing one sided posteriors, which could be mistakenly interpreted as bias. This leaves us with 1,986 samples in our test set. Fig.  \ref{fig:emu_vs_sim} shows a comparison of the predictions given by the emulator and the simulator. We find good agreement for the virial temperature and the ionization efficiency. The mean free path of ionizing photons shows the highest disagreement. This is likely because it is the hardest to constrain, since we have qualitatively observed that it has the least impact on the 2DPS. $\mathrm{R_{mfp}}$ only impacts the 2DPS at the late stages of ionization, where the radius of ionized bubbles approach the  $\mathrm{R_{mfp}}$ \citep{greig2017simultaneously}. Furthermore, \cite{greig2017simultaneously} find that values > 15 Mpc have little impact on the 21 cm power spectra, as clustering of ionizing sources becomes the dominant source of power, meaning we should expect wide constraints for high values of $\mathrm{R_{mfp}}$.

\subsection{Posterior Coverage Tests}\label{}
\subsubsection{Simulation Based Calibration} \label{SBC}
In order to assess the performance of our inference method, we implement a test known as Simulation Based Calibration (SBC), as seen in \citep{talts2018validating}. This test has been applied to other 21 cm inference pipelines e.g. \citep{meriot2025comparison, prelogovic2023exploring}. For each noisy simulation in our test set, we perform inference to obtain N samples (1,000 samples). For the case of NPE, we can sample directly from our modeled posterior without requiring the use of MCMC.

We then calculate the rank statistic for each parameter using the following:

\begin{equation}
r = \sum^{N}_{n=1} \mathbf{1}\left[\theta_{n}^{\mathrm{sample}} < \theta^{\mathrm{true}}\right],
\end{equation}

where $\theta_{n}^{\mathrm{sample}}$ is the $n^{\mathrm{th}}$ sample of the posterior distribution, and $\theta^{\mathrm{true}}$ is the value of the astrophysical parameter used to generate the mock observation. The rank statistic is simply the number of samples which fall below the true value. 

In the case of a perfectly calibrated posterior, we expect flat histograms. If the posterior is overconfident, we expect the true parameters to fall on the edges of our posterior more often than they should, giving us a U-shaped histogram. Conversely, if they are under-confident, we expect a hump shape, as the true parameter falls in the center too often. If the posterior is biased, we expect the true parameter to fall on one side more often than the other, and this bias is also reflected in the histograms \citep{talts2018validating}.

\begin{figure}
	\includegraphics[width=1\columnwidth]{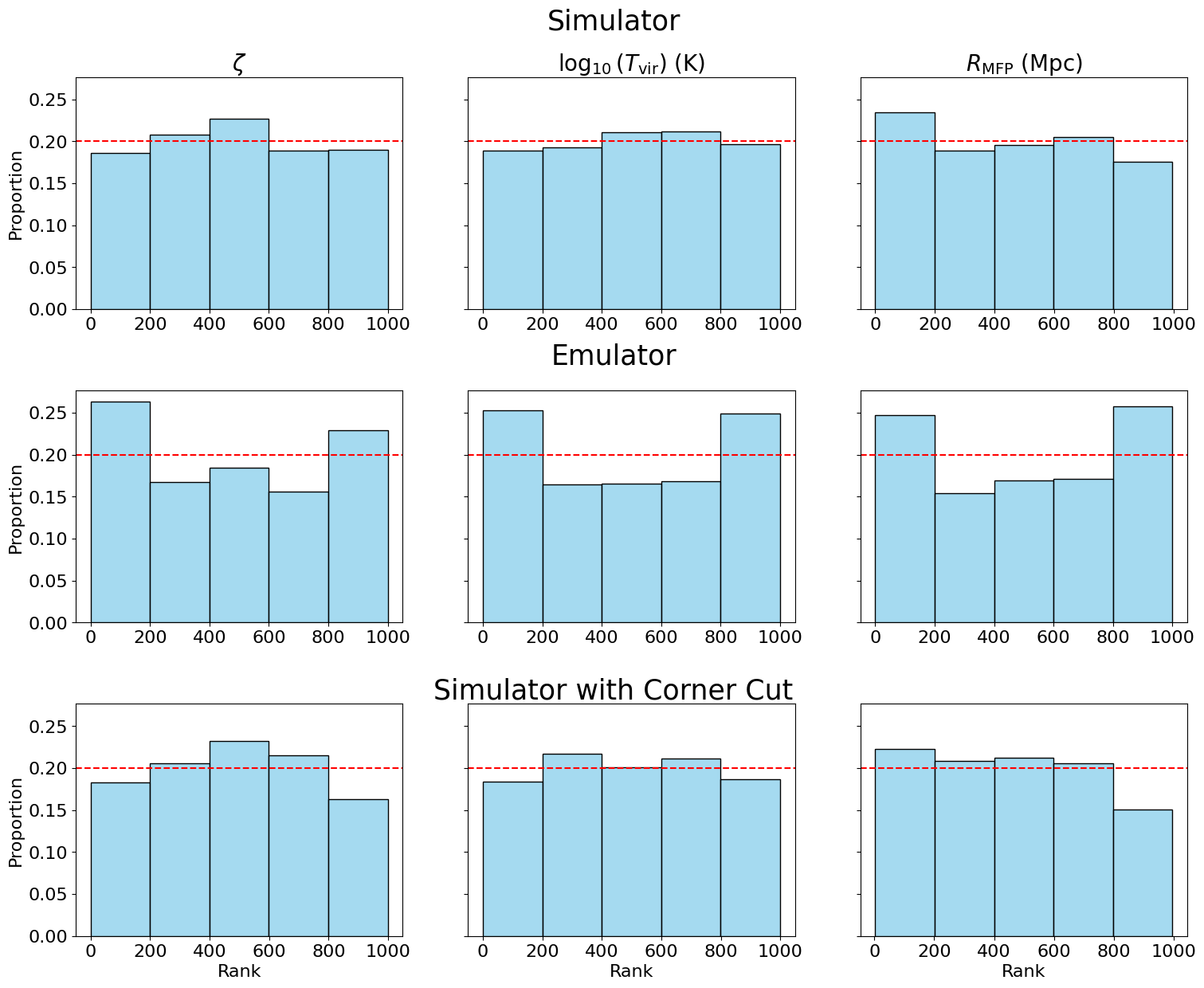}
    \caption{Simulation Based Calibration for the Neural Posterior Estimator trained on simulations, emulator, and simulations with the corner removed. Each histogram represents one of the three target astrophysical parameters, obtained using 1,000 samples for 1,986 test inferences. The red dashed line indicates perfect calibration.}
    \label{fig:sbc} 
\end{figure}

The simulation based calibration for our neural posterior estimator is shown in Fig.  \ref{fig:sbc}. In the case of the pipeline trained on simulations, the ionization efficiency and virial temperature parameters display some slight underconfidence. There is a slight bias in the mean free path of ionizing radiation, which is expected as it has been recovered the least successfully, since it has the least impact on the 2DPS. 

The simulation based calibration in the case of the emulator demonstrates slight overconfidence in all three parameters. We argue that underconfidence is preferred to slight overconfidence, as it is worse to rule out parts of parameter space where the true value might lie than to assign some probability to parts of parameter space where the true value doesn't fall. This indicates that increased sample density when using the emulator doesn't improve constraints.
We want to determine whether this is due to model error or loss of information when removing the corner. Hence, we also investigate the SBC for the pipeline trained on simulations with the corner removed. Here, we notice that the pipeline trained on simulations with the corner cut performs similarly to the full simulations, with both displaying slight underconfidence. This indicates model error in the emulator is a contributing factor to the poorer performance of the pipeline trained on emulated samples. This could be due to the mis-modeling of the stochasticity of the emulator due to sample variance.

\subsubsection{TARP}
\begin{figure}

	\includegraphics[width=1\columnwidth]{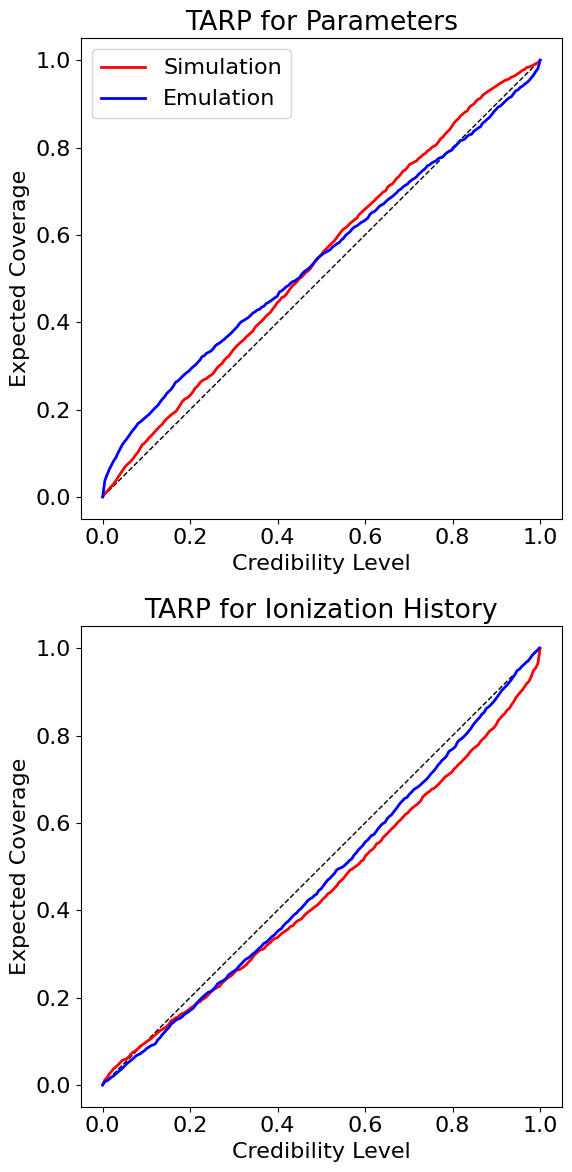}
    \caption{Posterior Coverage for the astrophysical parameters and ionization histories.  The lines indicate the posterior coverage for 1,986 test samples. The red lines are the posterior coverage for the SBI pipeline trained on simulations. The blue lines are the posterior coverage for the pipeline trained on emulated parameters. The dashed line is perfect calibration.}
    \label{fig:tarp} 
\end{figure}

Posterior calibration can also be assessed via the Test of Accuracy to Random Points (TARP) \citep{lemos2023sampling}. This method assesses the performance of the n-dimensional posterior, rather than the 1D marginalized posteriors, as in SBC. For a given set of data, the TARP test works by obtaining samples from the posterior distribution. A reference point is then drawn at a random point in parameter space. A sphere is drawn, centered on the true parameter value, with radius equal to the distance between the true parameter value and the reference point. The fraction of samples, f, that fall within the ball is then calculated. If the posterior is well calibrated, we expect f to be uniformly distributed. For further details, see \citep{lemos2023sampling}. For a well-calibrated posterior, we expect the expected coverage to equal the credibility level.

 The posterior calibration over the astrophysical parameters and the neutral fraction is shown in Fig.  \ref{fig:tarp} for both of the pipelines. Our predicted percentiles closely match the ideal case for both the pipeline trained on simulations and the pipeline trained on emulated samples. This indicates that across three dimensions, our inference is neither overly confident nor conservative. There is slight bias seen in both cases. The predicted percentiles for the astrophysical parameters are closer to the ideal case in the case of the simulations, again supporting that the emulator does not help improve predictions. There is no indication of overconfidence or of being underconfident in either case, however, there is slight bias in both cases. This supports that both pipelines are able to accurately recover the ionization history. 

\section{Discussion}\label{discussion}

In this paper we have presented a method for inferring astrophysical parameters of interest and the ionization history from noisy 21cm 2DPS mock observations. We have contrasted training the inference method on 21cmFAST simulations and samples obtained with an emulator of the 2DPS. We find both methods are able to recover the ionization history with high accuracy. However, the pipeline displays a preference towards simulated training samples.

Our development of an emulator of the 2D power spectrum is a relatively novel task. While emulators of the 1D power spectrum have been explored in works like \cite{Schmit_2017} and \cite{Breitman_2023}, a widely used method for creating a 2DPS emulator has yet to be decided. Possible obstacles involved with building 2DPS emulators include dealing with sample variance and creating expensive simulation banks. With the recent conclusion of the SKA Science Data Challenge 3b, an inference task focused on retrieving the ionization history associated with the simulated observations of the 2DPS, a multitude of 2DPS emulator model architectures are expected to be produced, which will provide robust comparison between methods and allow the 21 cm community to agree on a best emulator implementation moving forward.

When designing our emulator, the effect of sample variance at low $k$ scales meant that our emulator would be unable to predict accurately at the lowest scales due to the stochasticity present. Removing three of the pixels to improve the performance of the emulator is one method of dealing with the sample variance, which allows us to regain some accuracy in the inference step. The error of an emulator trained on all $k$-bins was much higher than our emulator trained with the three corner $k$ pixels masked. 
Another method to account for the SV, is a non-deterministic network, predicting from a learned distribution of the sample variance. We mirrored this effect by reintroducing an approximation of the sample variance in the SBI step to take into account the stochasticity. However, as discussed in Section \ref{sec:inference}, this may not be the most appropriate model, as it assumes the sample variance is centered on the emulator prediction. As we had limited realizations of the 2DPS per parameter combination in our training set, this assumption may not be accurate. Furthermore, modeling the sample variance this way is equivalent to a Gaussian approximation,  which is inaccurate for reasons discussed in Section \ref{SBI}.

We were able to achieve most errors under or around the 10\% level, however, introducing a data compression method for the training of the emulator may have improved its performance. Our input dimensions for the emulator were only 3, while the output was 488. Reducing the dimensionality of the output whilst minimizing information loss in the process is a valuable analysis technique. Using principal component analysis (PCA) or a neural network designed for dimensionality reduction are possible routes to explore before training the emulator. We leave this application to future work. 

We consider two sources for the pipeline's improved performance when trained on the simulations. The first source is loss of information when removing the corner $k$ pixels, and the second source is model error within the emulator itself. We explored retraining the pipeline on simulations with the same pixel masking in order to determine the dominant source of error. Furthermore, as shown in the SBC tests in Fig.  \ref{fig:sbc}, the simulations with the corner cut perform comparably to the full simulations, indicating emulator model error is the largest contributor to the decreased performance.
Both pipelines have been able to recover the virial temperature and ionizing efficiency with high accuracy. Our method demonstrated better performance when trained on simulated data than emulated samples. This is particularly true for the recovery of the mean free path of ionizing radiation, as seen in Fig. \ref{fig:sbc} indicating that use of an emulator of the 2DPS has not added benefit.

Our work indicates that due to the complexity of the 2DPS, emulators of the 1DPS are a more robust choice when performing inference, such as in \cite{meriot2025comparison}, when considering simple dense models. There are two factors which may be contributing to this effect. The first is that the modeling of the sample variance may not be sufficient, or that the emulator itself is contributing non-negligible model error. For this reason, more complex models where the stochasticity of the 2DPS can be better accounted for should be explored. SBI with simulations was able to provide accurate constraints on both the IGM properties, and the ionization history, indicating an emulator may not be needed for this analysis.

We have demonstrated that in the cases of a late reionization we obtain tight constraints on astrophysical parameters and ionization history. Our constraints are much wider for the case of early reionization, as there is much less information contained within the observation. This is due to a combination of the noise dependence on frequency, and the number of bins with zero data after reionization. However, we note that measurements of the CMB indicate a reionization midpoint at around redshift $z = 7.7$ \citep{aghanim2020planck}, hence early reionization scenarios are less physically motivated. Likely reionization scenarios are therefore constrained well. 

Both inference training sets accurately recovered the ionization history. Despite a slightly improved performance for the pipeline trained on simulations, we have demonstrated that the 2DPS emulator is able to recover similar constraints. This emulator has the advantage of enabling flexible inference as, once trained, it does not rely on an amortized bank of simulations.  This would allow exploration of MCMC or sequential neural posterior estimation (SNPE) techniques  \citep{papamakarios2019sequential}. This type of learning is impossible with a pre-run bank of simulations, highlighting the potential use of the developed emulator. 

This inference method is limited by its reliance on the model used to generate the simulated data. A big problem facing the 21 cm community is model misspecification. Semi-numerical codes and full radiative transfer codes can behave quite differently, meaning performing inference on a mock observation generated with a different code from the training might fail. We acknowledge that our pipeline might only lead to accurate constraints with an observation generated with 21cmFAST, which is a problem the community will face when performing inference on a real 2DPS detection, and exploring the significance of this effect is one of the goals of the SKA Data Challenge 3b. 

Furthermore, this work has assumed perfect foreground removal and no instrumental systematics. This is optimistic regarding future observations. In an SBI pipeline, these effects are more straight forward to incorporate than within an MCMC framework. Instrumental effects such as beam smoothing, or foreground residuals could have been applied to our lightcones as an additional processing step before calculating the power spectra. However, in practice, simulating the foreground residuals would add to the computational cost and are beyond the scope of this work. Removing the foregrounds was the target of the SKA SDC3a \citep{bonaldi2025square}, in which some teams demonstrated accurate foreground removal. An alternative strategy is to remove the wedge of the power spectra which is most affected by foregrounds, as demonstrated in \cite{greig2024inferring}, this leads to loss of constraining power. Inference including foregrounds was done by \cite{zhao2024simulationbasedinferencereionizationparameters}, where they concluded that inference with foreground residuals resulted in larger uncertainties on their posteriors. We expect the addition of foregrounds would have had a similar effect on our inference pipeline. We defer inclusion of instrumental systematics and foreground residuals to future work.

Our inference method uses information held only within the 2DPS. However, the SKA aims to make full brightness temperature intensity maps. The 21 cm field contains non-Gaussian information, which is thrown away when taking the power spectra. \cite{prelogovic2024informative} have demonstrated the additional information contained within the field. Exploring other summary statistics could help improve constraints, which we plan to investigate in future research.

\section{Conclusions}

In this work, we have contrasted two training sets for simulation based inference from the 21 cm signal. We inferred the astrophysical parameters of our model, 21cmFAST, and the ionization history between redshift z $\sim$ 5 and  z $\sim$ 12.  We compared training our neural density estimator on noisy 21cmFAST simulations, and noisy samples generated from an emulator of the 2DPS. We developed an accurate emulator of the 2DPS in eight redshift bins. Due to the effect of sample variance, we mask the three corner $k$- mode pixels most affected by the SV. The model's prediction error is then at or under 10\% of the spread of the noise and sample variance.

We train our SBI method to recover the virial temperature, $T_{\mathrm{vir}}$, the ionizing efficiency, $\zeta$, and the mean free path of ionizing photons, $R_{\mathrm{mfp}}$. We demonstrate that both methods are able to accurately recover the virial temperature and the ionizing efficiency, with a degeneracy in these two parameters. The inference method recovers $R_{\mathrm{mfp}}$ with wider constraints.  The pipeline shows a preference towards training on simulations, as demonstrated in simulation based calibration tests in Section \ref{SBC}. We attribute this to the emulator being unable to correctly capture the effects of sample variance. We, therefore, conclude that this work suggests that for the 2DPS, inference may be more robust when performed on simulations rather than emulator of the 2DPS. It is likely this performance is dependent on the choice of 21 cm simulation code, and our pipeline is likely to only lead to accurate constraints when performing inference on a mock observation generated with 21cmFAST.

We also trained an emulator from these astrophysical parameters to the ionization history. We can then sample from our learnt posteriors. We then evaluate our ionization history emulator at these points in parameter space to obtain samples from the posterior distribution of the ionization history at 41 redshifts. Both pipelines are able to recover the ionization history with extremely high accuracy, and neither method is strongly preferred. Since the 2DPS emulator is able to recover the ionization history with comparable accuracy, we conclude that it provides the benefit of more flexible inference for future inference. 

\section*{Acknowledgements}

NC and CN thank the Imperial Centre for Doctoral Training in Data Intensive Physics for funding this work. JRP and RM acknowledge support from STFC grant ST/Y004132/1. The results in this paper have been derived using the {\small SBI}, {\small LTU-ILI}, {\small 21cmFAST}, {\small tools21cm} and {\small PyTorch} packages. We wish to thank T. Lucas Makinen for useful discussions. 

%%%%%%%%%%%%%%%%%%%%%%%%%%%%%%%%%%%%%%%%%%%%%%%%%%
\section*{Data Availability}

Our code is available upon request via email. The data underlying this article are available from the authors on
reasonable request.



\bibliographystyle{mnras}
\bibliography{ref} % if your bibtex file is called example.bib

@article{field1958excitation,
  title={Excitation of the hydrogen 21-cm line},
  author={Field, George B},
  journal={Proceedings of the IRE},
  volume={46},
  number={1},
  pages={240--250},
  year={1958},
  publisher={IEEE}
}

@article{pritchard201221,
  title={21 cm cosmology in the 21st century},
  author={Pritchard, Jonathan R and Loeb, Abraham},
  journal={Reports on Progress in Physics},
  volume={75},
  number={8},
  pages={086901},
  year={2012},
  publisher={IOP Publishing}}

@article{liu2020data,
  title={Data analysis for precision 21 cm cosmology},
  author={Liu, Adrian and Shaw, J Richard},
  journal={PASP},
  volume={132},
  number={1012},
  pages={062001},
  year={2020},
  publisher={IOP Publishing}
}

@article{wouthuysen1952excitation,
  title={On the excitation mechanism of the 21-cm (radio-frequency) interstellar hydrogen emission line.},
  author={Wouthuysen, SA},
  journal={AJ},
  volume={57},
  pages={31--32},
  year={1952}
}

@article{paciga2013simulation,
  title={A simulation-calibrated limit on the H i power spectrum from the GMRT Epoch of Reionization experiment},
  author={Paciga, Gregory and Albert, Joshua G and Bandura, Kevin and Chang, Tzu-Ching and Gupta, Yashwant and Hirata, Christopher and Odegova, Julia and Pen, Ue-Li and Peterson, Jeffrey B and Roy, Jayanta and others},
  journal={MNRAS},
  volume={433},
  number={1},
  pages={639--647},
  year={2013},
  publisher={The Royal Astronomical Society}
}

@article{mertens2020improved,
  title={Improved upper limits on the 21 cm signal power spectrum of neutral hydrogen at z≈ 9.1 from LOFAR},
  author={Mertens, Florent G and Mevius, Maaijke and Koopmans, Leon VE and Offringa, AR and Mellema, Garrelt and Zaroubi, Saleem and Brentjens, MA and Gan, H and Gehlot, Bharat Kumar and Pandey, VN and others},
  journal={MNRAS},
  volume={493},
  number={2},
  pages={1662--1685},
  year={2020},
  publisher={Oxford University Press}
}

@article{ghara2020constraining,
  title={Constraining the intergalactic medium at z≈ 9.1 using LOFAR Epoch of Reionization observations},
  author={Ghara, Raghunath and Giri, Sambit Kumar and Mellema, Garrelt and Ciardi, Benedetta and Zaroubi, Saleem and Iliev, Ilian T and Koopmans, Leon VE and Chapman, Emma and Gazagnes, S and Gehlot, Bharat Kumar and others},
  journal={MNRAS},
  volume={493},
  number={4},
  pages={4728--4747},
  year={2020},
  publisher={Oxford University Press}
}

@article{ceccotti2025first,
  title={First upper limits on the 21-cm signal power spectrum of neutral hydrogen at $ z= 9.16$ from the LOFAR 3C196 field},
  author={Ceccotti, E and Offringa, AR and Mertens, FG and Koopmans, LVE and Munshi, S and Chege, JK and Acharya, A and Brackenhoff, SA and Chapman, E and Ciardi, B and others},
  journal={arXiv preprint arXiv:2504.18534},
  year={2025}
}

@article{deboer2017hydrogen,
  title={Hydrogen epoch of reionization array (HERA)},
  author={DeBoer, David R and Parsons, Aaron R and Aguirre, James E and Alexander, Paul and Ali, Zaki S and Beardsley, Adam P and Bernardi, Gianni and Bowman, Judd D and Bradley, Richard F and Carilli, Chris L and others},
  journal={PASP},
  volume={129},
  number={974},
  pages={045001},
  year={2017},
  publisher={IOP Publishing}
}

@article{kolopanis2023new,
  title={New EoR power spectrum limits from MWA Phase II using the delay spectrum method and novel systematic rejection},
  author={Kolopanis, Matthew and Pober, Jonathan C and Jacobs, Daniel C and McGraw, Samantha},
  journal={MNRAS},
  volume={521},
  number={4},
  pages={5120--5138},
  year={2023},
  publisher={Oxford University Press}
}

@article{li2019first,
  title={First season MWA Phase II epoch of reionization power spectrum results at redshift 7},
  author={Li, W and Pober, JC and Barry, N and Hazelton, BJ and Morales, MF and Trott, CM and Lanman, A and Wilensky, M and Sullivan, I and Beardsley, AP and others},
  journal={ApJ},
  volume={887},
  number={2},
  pages={141},
  year={2019},
  publisher={IOP Publishing}
}

@article{barry2019improving,
  title={Improving the epoch of reionization power spectrum results from Murchison widefield array season 1 observations},
  author={Barry, N and Wilensky, M and Trott, CM and Pindor, B and Beardsley, AP and Hazelton, BJ and Sullivan, IS and Morales, MF and Pober, JC and Line, J and others},
  journal={ApJ},
  volume={884},
  number={1},
  pages={1},
  year={2019},
  publisher={IOP Publishing}
}

@article{ho2024ltu,
  title={Ltu-ili: An all-in-one framework for implicit inference in astrophysics and cosmology},
  author={Ho, Matthew and Bartlett, Deaglan J and Chartier, Nicolas and Cuesta-Lazaro, Carolina and Ding, Simon and Lapel, Axel and Lemos, Pablo and Lovell, Christopher C and Makinen, T Lucas and Modi, Chirag and others},
  journal={arXiv preprint arXiv:2402.05137},
  year={2024}
}

@article{BoeltsDeistler_sbi_2025,
  doi = {10.21105/joss.07754},
  url = {https://doi.org/10.21105/joss.07754},
  year = {2025},
  publisher = {The Open Journal},
  volume = {10},
  number = {108},
  pages = {7754},
  author = {Jan Boelts and Michael Deistler and Manuel Gloeckler and Álvaro Tejero-Cantero and Jan-Matthis Lueckmann and Guy Moss and Peter Steinbach and Thomas Moreau and Fabio Muratore and Julia Linhart and Conor Durkan and Julius Vetter and Benjamin Kurt Miller and Maternus Herold and Abolfazl Ziaeemehr and Matthijs Pals and Theo Gruner and Sebastian Bischoff and Nastya Krouglova and Richard Gao and Janne K. Lappalainen and Bálint Mucsányi and Felix Pei and Auguste Schulz and Zinovia Stefanidi and Pedro Rodrigues and Cornelius Schröder and Faried Abu Zaid and Jonas Beck and Jaivardhan Kapoor and David S. Greenberg and Pedro J. Gonçalves and Jakob H. Macke},
  title = {sbi reloaded: a toolkit for simulation-based inference workflows},
  journal = {Journal of Open Source Software}
}

@article{papamakarios2019neural,
  title={Neural density estimation and likelihood-free inference},
  author={Papamakarios, George},
  journal={arXiv preprint arXiv:1910.13233},
  year={2019}
}

@article{bishop1994mixture,
  title={Mixture density networks (WorkingPaper)},
  author={Bishop, C},
  journal={Aston University, Birmingham, UK},
  year={1994}
}

@article{papamakarios2021normalizing,
  title={Normalizing flows for probabilistic modeling and inference},
  author={Papamakarios, George and Nalisnick, Eric and Rezende, Danilo Jimenez and Mohamed, Shakir and Lakshminarayanan, Balaji},
  journal={Journal of Machine Learning Research},
  volume={22},
  number={57},
  pages={1--64},
  year={2021}
}

@article{durkan2019neural,
  title={Neural spline flows},
  author={Durkan, Conor and Bekasov, Artur and Murray, Iain and Papamakarios, George},
  journal={Advances in neural information processing systems},
  volume={32},
  year={2019}
}

@article{talts2018validating,
  title={Validating Bayesian inference algorithms with simulation-based calibration},
  author={Talts, Sean and Betancourt, Michael and Simpson, Daniel and Vehtari, Aki and Gelman, Andrew},
  journal={arXiv preprint arXiv:1804.06788},
  year={2018}
}

@article{Breitman_2023,
   title={<scp>21cmemu</scp>: an emulator of <scp>21cmfast</scp> summary observables},
   volume={527},
   ISSN={1365-2966},
   url={http://dx.doi.org/10.1093/mnras/stad3849},
   DOI={10.1093/mnras/stad3849},
   number={4},
   journal={MNRAS},
   publisher={Oxford University Press (OUP)},
   author={Breitman, Daniela and Mesinger, Andrei and Murray, Steven G and Prelogović, David and Qin, Yuxiang and Trotta, Roberto},
   year={2023},
   month=dec, pages={9833–9852} }

@article{Bye_2022,
   title={21cmVAE: A Very Accurate Emulator of the 21 cm Global Signal},
   volume={930},
   ISSN={1538-4357},
   url={http://dx.doi.org/10.3847/1538-4357/ac6424},
   DOI={10.3847/1538-4357/ac6424},
   number={1},
   journal={ApJ},
   publisher={American Astronomical Society},
   author={Bye, Christian H. and Portillo, Stephen K. N. and Fialkov, Anastasia},
   year={2022},
   month=may, pages={79} }

@article{koopmans2015cosmic,
  title={The cosmic dawn and epoch of reionization with the square kilometre array},
  author={Koopmans, LVE and Pritchard, Jonathan and Mellema, G and Abdalla, F and Aguirre, J and Ahn, K and Barkana, R and Van Bemmel, I and Bernardi, G and Bonaldi, A and others},
  journal={arXiv preprint arXiv:1505.07568},
  year={2015}
}

@article{greig2024inferring,
  title={Inferring astrophysical parameters using the 2D cylindrical power spectrum from reionization},
  author={Greig, Bradley and Prelogovi{\'c}, David and Qin, Yuxiang and Ting, Yuan-Sen and Mesinger, Andrei},
  journal={MNRAS},
  volume={533},
  number={2},
  pages={2530--2545},
  year={2024},
  publisher={Oxford University Press}
}

@article{meriot2025comparison,
       author = {{Meriot}, R. and {Semelin}, B. and {Cornu}, D.},
        title = "{Comparison of Bayesian inference methods using the LORELI II database of hydro-radiative simulations of the 21-cm signal}",
      journal = {\aap},
         year = 2025,
        month = jun,
       volume = {698},
          eid = {A80},
        pages = {A80},
          doi = {10.1051/0004-6361/202452901},
archivePrefix = {arXiv},
       eprint = {2411.03093},
 primaryClass = {astro-ph.CO},
       adsurl = {https://ui.adsabs.harvard.edu/abs/2025A&A...698A..80M}
}

@article{saxena2024simulation,
  title={Simulation-based inference of the sky-averaged 21-cm signal from CD-EoR with REACH},
  author={Saxena, Anchal and Meerburg, P Daniel and Weniger, Christoph and Acedo, Eloy de Lera and Handley, Will},
  journal={RAS Techniques and Instruments},
  volume={3},
  number={1},
  pages={724--736},
  year={2024},
  publisher={Oxford University Press}
}

@article{Kern_2017,
   title={Emulating Simulations of Cosmic Dawn for 21 cm Power Spectrum Constraints on Cosmology, Reionization, and X-Ray Heating},
   volume={848},
   ISSN={1538-4357},
   url={http://dx.doi.org/10.3847/1538-4357/aa8bb4},
   DOI={10.3847/1538-4357/aa8bb4},
   number={1},
   journal={ApJ},
   publisher={American Astronomical Society},
   author={Kern, Nicholas S. and Liu, Adrian and Parsons, Aaron R. and Mesinger, Andrei and Greig, Bradley},
   year={2017},
   month=oct, pages={23} }

@article{Schmit_2017,
   title={Emulation of reionization simulations for Bayesian inference of astrophysics parameters using neural networks},
   volume={475},
   ISSN={1365-2966},
   url={http://dx.doi.org/10.1093/mnras/stx3292},
   DOI={10.1093/mnras/stx3292},
   number={1},
   journal={MNRAS},
   publisher={Oxford University Press (OUP)},
   author={Schmit, C J and Pritchard, J R},
   year={2017},
   month=dec, pages={1213–1223} }

@article{Nadler_2025,
   title={The Impact of Molecular Hydrogen Cooling on the Galaxy Formation Threshold},
   volume={983},
   ISSN={2041-8213},
   url={http://dx.doi.org/10.3847/2041-8213/adbc6e},
   DOI={10.3847/2041-8213/adbc6e},
   number={1},
   journal={ApJL},
   publisher={American Astronomical Society},
   author={Nadler, Ethan O.},
   year={2025},
   month=apr, pages={L23} }

@inproceedings{lemos2023sampling,
  title={Sampling-based accuracy testing of posterior estimators for general inference},
  author={Lemos, Pablo and Coogan, Adam and Hezaveh, Yashar and Perreault-Levasseur, Laurence},
  booktitle={International Conference on Machine Learning},
  pages={19256--19273},
  year={2023},
  organization={PMLR}
}

@article{giri2020tools21cm,
  title={Tools21cm: A python package to analyse the large-scale 21-cm signal from the Epoch of Reionization and Cosmic Dawn},
  author={Giri, Sambit and Mellema, Garrelt and Jensen, Hannes},
  journal={Journal of Open Source Software},
  volume={5},
  number={52},
  pages={2363},
  year={2020},
  publisher={Open Journals}
}

@article{mcquinn2006cosmological,
  title={Cosmological parameter estimation using 21 cm radiation from the epoch of reionization},
  author={McQuinn, Matthew and Zahn, Oliver and Zaldarriaga, Matias and Hernquist, Lars and Furlanetto, Steven R},
  journal={ApJ},
  volume={653},
  number={2},
  pages={815},
  year={2006},
  publisher={IOP Publishing}
}

@article{greig201521cmmc,
  title={21CMMC: an MCMC analysis tool enabling astrophysical parameter studies of the cosmic 21 cm signal},
  author={Greig, Bradley and Mesinger, Andrei},
  journal={MNRAS},
  volume={449},
  number={4},
  pages={4246--4263},
  year={2015},
  publisher={The Royal Astronomical Society}
}

@article{mellema2006c2,
  title={C2-ray: A new method for photon-conserving transport of ionizing radiation},
  author={Mellema, Garrelt and Iliev, Ilian T and Alvarez, Marcelo A and Shapiro, Paul R},
  journal={New Astronomy},
  volume={11},
  number={5},
  pages={374--395},
  year={2006},
  publisher={Elsevier}
}

@article{semelin201721ssd,
  title={21SSD: a public data base of simulated 21-cm signals from the epoch of reionization},
  author={Semelin, Benoit and Eames, Evan and Bolgar, Florian and Caillat, Michel},
  journal={MNRAS},
  volume={472},
  number={4},
  pages={4508--4520},
  year={2017},
  publisher={Oxford University Press}
}

@article{ocvirk2020cosmic,
  title={Cosmic Dawn II (CoDa II): a new radiation-hydrodynamics simulation of the self-consistent coupling of galaxy formation and reionization},
  author={Ocvirk, Pierre and Aubert, Dominique and Sorce, Jenny G and Shapiro, Paul R and Deparis, Nicolas and Dawoodbhoy, Taha and Lewis, Joseph and Teyssier, Romain and Yepes, Gustavo and Gottl{\"o}ber, Stefan and others},
  journal={MNRAS},
  volume={496},
  number={4},
  pages={4087--4107},
  year={2020},
  publisher={Oxford University Press}
}

@article{murray202021cmfast,
  title={21cmFAST v3: A Python-integrated C code for generating 3D realizations of the cosmic 21cm signal},
  author={Murray, Steven G and Greig, Bradley and Mesinger, Andrei and Mu{\~n}oz, Julian B and Qin, Yuxiang and Park, Jaehong and Watkinson, Catherine A},
  journal={arXiv preprint arXiv:2010.15121},
  year={2020}
}

@article{mesinger201121cmfast,
  title={21CMFAST: a fast, seminumerical simulation of the high-redshift 21-cm signal},
  author={Mesinger, Andrei and Furlanetto, Steven and Cen, Renyue},
  journal={MNRAS},
  volume={411},
  number={2},
  pages={955--972},
  year={2011},
  publisher={Blackwell Publishing Ltd Oxford, UK}
}

@article{aghanim2020planck,
  title={Planck 2018 results-VI. Cosmological parameters},
  author={Aghanim, Nabila and Akrami, Yashar and Ashdown, Mark and Aumont, Jonathan and Baccigalupi, Carlo and Ballardini, Mario and Banday, Anthony J and Barreiro, RB and Bartolo, Nicola and Basak, S and others},
  journal={A\&A},
  volume={641},
  pages={A6},
  year={2020},
  publisher={EDP sciences}
}

@article{cranmer2020frontier,
  title={The frontier of simulation-based inference},
  author={Cranmer, Kyle and Brehmer, Johann and Louppe, Gilles},
  journal={Proceedings of the National Academy of Sciences},
  volume={117},
  number={48},
  pages={30055--30062},
  year={2020},
  publisher={National Academy of Sciences}
}

@article{papamakarios2016fast,
  title={Fast $\varepsilon$-free inference of simulation models with bayesian conditional density estimation},
  author={Papamakarios, George and Murray, Iain},
  journal={Advances in neural information processing systems},
  volume={29},
  year={2016}
}

@article{furlanetto2006cosmology,
  title={Cosmology at low frequencies: The 21 cm transition and the high-redshift Universe},
  author={Furlanetto, Steven R and Oh, S Peng and Briggs, Frank H},
  journal={Physics Reports},
  volume={433},
  number={4-6},
  pages={181--301},
  year={2006},
  publisher={Elsevier}
}

@article{loeb2004measuring,
  title={Measuring the Small-Scale Power Spectrum of Cosmic Density Fluctuations<? format?> through 21 cm Tomography Prior to the Epoch of Structure Formation},
  author={Loeb, Abraham and Zaldarriaga, Matias},
  journal={Physical Review Letters},
  volume={92},
  number={21},
  pages={211301},
  year={2004},
  publisher={APS}
}

@article{hirata2006wouthuysen,
  title={Wouthuysen-Field coupling strength and application to high-redshift 21-cm radiation},
  author={Hirata, Christopher M},
  journal={MNRAS},
  volume={367},
  number={1},
  pages={259--274},
  year={2006},
  publisher={Blackwell Science Ltd 23 Ainslie Place, Edinburgh EH3 6AJ, UK. Telephone~…}
}

@article{loeb2001reionization,
  title={The reionization of the universe by the first stars and quasars},
  author={Loeb, Abraham and Barkana, Rennan},
  journal={Annual review of A\&A},
  volume={39},
  number={1},
  pages={19--66},
  year={2001},
  publisher={Annual Reviews 4139 El Camino Way, PO Box 10139, Palo Alto, CA 94303-0139, USA}
}

@article{tejero-cantero2020sbi,
  doi = {10.21105/joss.02505},
  url = {https://doi.org/10.21105/joss.02505},
  year = {2020},
  publisher = {The Open Journal},
  volume = {5},
  number = {52},
  pages = {2505},
  author = {Alvaro Tejero-Cantero and Jan Boelts and Michael Deistler and Jan-Matthis Lueckmann and Conor Durkan and Pedro J. Gonçalves and David S. Greenberg and Jakob H. Macke},
  title = {sbi: A toolkit for simulation-based inference},
  journal = {Journal of Open Source Software}
}

@article{lueckmann2017flexible,
  title={Flexible statistical inference for mechanistic models of neural dynamics},
  author={Lueckmann, Jan-Matthis and Goncalves, Pedro J and Bassetto, Giacomo and {\"O}cal, Kaan and Nonnenmacher, Marcel and Macke, Jakob H},
  journal={Advances in neural information processing systems},
  volume={30},
  year={2017}
}

@inproceedings{papamakarios2019sequential,
  title={Sequential neural likelihood: Fast likelihood-free inference with autoregressive flows},
  author={Papamakarios, George and Sterratt, David and Murray, Iain},
  booktitle={The 22nd international conference on artificial intelligence and statistics},
  pages={837--848},
  year={2019},
  organization={PMLR}
}

@article{prelogovic2024informative,
  title={How informative are summaries of the cosmic 21 cm signal?},
  author={Prelogovi{\'c}, David and Mesinger, Andrei},
  journal={A\&A},
  volume={688},
  pages={A199},
  year={2024},
  publisher={EDP Sciences}
}

@article{de2021inference,
  title={Inference of the optical depth to reionization from low multipole temperature and polarization Planck data},
  author={de Belsunce, Roger and Gratton, Steven and Coulton, William and Efstathiou, George},
  journal={MNRAS},
  volume={507},
  number={1},
  pages={1072--1091},
  year={2021},
  publisher={Oxford University Press}
}

@article{heinrich2021reionization,
  title={Reionization effective likelihood from Planck 2018 data},
  author={Heinrich, Chen and Hu, Wayne},
  journal={Physical Review D},
  volume={104},
  number={6},
  pages={063505},
  year={2021},
  publisher={APS}
}

@article{fan2006constraining,
  title={Constraining the evolution of the ionizing background and the epoch of reionization with z~ 6 quasars. II. A sample of 19 quasars},
  author={Fan, Xiaohui and Strauss, Michael A and Becker, Robert H and White, Richard L and Gunn, James E and Knapp, Gillian R and Richards, Gordon T and Schneider, Donald P and Brinkmann, J and Fukugita, Masataka},
  journal={AJ},
  volume={132},
  number={1},
  pages={117},
  year={2006},
  publisher={IOP Publishing}
}

@article{becker2007evolution,
  title={The Evolution of Optical Depth in the Ly$\alpha$ Forest: Evidence Against Reionization at z\~{} 6},
  author={Becker, George D and Rauch, Michael and Sargent, Wallace LW},
  journal={ApJ},
  volume={662},
  number={1},
  pages={72},
  year={2007},
  publisher={IOP Publishing}
}

@article{bosman2018new,
  title={New constraints on Lyman-$\alpha$ opacity with a sample of 62 quasarsat z> 5.7},
  author={Bosman, Sarah EI and Fan, Xiaohui and Jiang, Linhua and Reed, Sophie and Matsuoka, Yoshiki and Becker, George and Haehnelt, Martin},
  journal={MNRAS},
  volume={479},
  number={1},
  pages={1055--1076},
  year={2018},
  publisher={Oxford University Press}
}

@article{d2023xqr,
  title={XQR-30: The ultimate XSHOOTER quasar sample at the reionization epoch},
  author={D’Odorico, Valentina and Ba{\~n}ados, E and Becker, G\_D and Bischetti, Manuela and Bosman, S\_E\_I and Cupani, Guido and Davies, R and Farina, E\_P and Ferrara, Andrea and Feruglio, Chiara and others},
  journal={MNRAS},
  volume={523},
  number={1},
  pages={1399--1420},
  year={2023},
  publisher={Oxford University Press}
}

@article{bolton2011neutral,
  title={How neutral is the intergalactic medium surrounding the redshift z= 7.085 quasar ULAS J1120+ 0641?},
  author={Bolton, James S and Haehnelt, Martin G and Warren, Stephen J and Hewett, Paul C and Mortlock, Daniel J and Venemans, Bram P and McMahon, Richard G and Simpson, Chris},
  journal={MNRAS: Letters},
  volume={416},
  number={1},
  pages={L70--L74},
  year={2011},
  publisher={The Royal Astronomical Society}
}

@article{mortlock2011luminous,
  title={A luminous quasar at a redshift of z= 7.085},
  author={Mortlock, Daniel J and Warren, Stephen J and Venemans, Bram P and Patel, Mitesh and Hewett, Paul C and McMahon, Richard G and Simpson, Chris and Theuns, Tom and Gonz{\'a}les-Solares, Eduardo A and Adamson, Andy and others},
  journal={Nature},
  volume={474},
  number={7353},
  pages={616--619},
  year={2011},
  publisher={Nature Publishing Group UK London}
}

@article{banados2018800,
  title={An 800-million-solar-mass black hole in a significantly neutral Universe at a redshift of 7.5},
  author={Ba{\~n}ados, Eduardo and Venemans, Bram P and Mazzucchelli, Chiara and Farina, Emanuele P and Walter, Fabian and Wang, Feige and Decarli, Roberto and Stern, Daniel and Fan, Xiaohui and Davies, Frederick B and others},
  journal={Nature},
  volume={553},
  number={7689},
  pages={473--476},
  year={2018},
  publisher={Nature Publishing Group UK London}
}

@article{wang2020significantly,
  title={A Significantly Neutral Intergalactic Medium Around the Luminous z= 7 Quasar J0252--0503},
  author={Wang, Feige and Davies, Frederick B and Yang, Jinyi and Hennawi, Joseph F and Fan, Xiaohui and Barth, Aaron J and Jiang, Linhua and Wu, Xue-Bing and Mudd, Dale M and Ba{\~n}ados, Eduardo and others},
  journal={ApJ},
  volume={896},
  number={1},
  pages={23},
  year={2020},
  publisher={IOP Publishing}
}

@article{yang2020poniua,
  title={P{\=o}niu{\=a} ‘ena: A Luminous z= 7.5 Quasar Hosting a 1.5 Billion Solar Mass Black Hole},
  author={Yang, Jinyi and Wang, Feige and Fan, Xiaohui and Hennawi, Joseph F and Davies, Frederick B and Yue, Minghao and Banados, Eduardo and Wu, Xue-Bing and Venemans, Bram and Barth, Aaron J and others},
  journal={ApJL},
  volume={897},
  number={1},
  pages={L14},
  year={2020},
  publisher={IOP Publishing}
}

@article{ouchi2010statistics,
  title={Statistics of 207 Ly$\alpha$ emitters at a redshift near 7: constraints on reionization and galaxy formation models},
  author={Ouchi, Masami and Shimasaku, Kazuhiro and Furusawa, Hisanori and Saito, Tomoki and Yoshida, Makiko and Akiyama, Masayuki and Ono, Yoshiaki and Yamada, Toru and Ota, Kazuaki and Kashikawa, Nobunari and others},
  journal={ApJ},
  volume={723},
  number={1},
  pages={869},
  year={2010},
  publisher={IOP Publishing}
}

@article{clement2012evolution,
  title={Evolution of the observed Ly$\alpha$ luminosity function from z= 6.5 to z= 7.7: evidence for the epoch of reionization?},
  author={Cl{\'e}ment, B and Cuby, J-G and Courbin, F and Fontana, A and Freudling, W and Fynbo, J and Gallego, J and Hibon, P and Kneib, J-P and Le F{\`e}vre, O and others},
  journal={A \& A},
  volume={538},
  pages={A66},
  year={2012},
  publisher={EDP Sciences}
}

@article{konno2014accelerated,
  title={Accelerated evolution of the Ly$\alpha$ luminosity function at z≳ 7 revealed by the Subaru ultra-deep survey for Ly$\alpha$ emitters at z= 7.3},
  author={Konno, Akira and Ouchi, Masami and Ono, Yoshiaki and Shimasaku, Kazuhiro and Shibuya, Takatoshi and Furusawa, Hisanori and Nakajima, Kimihiko and Naito, Yoshiaki and Momose, Rieko and Yuma, Suraphong and others},
  journal={ApJ},
  volume={797},
  number={1},
  pages={16},
  year={2014},
  publisher={IOP Publishing}
}

@article{drake2017muse,
  title={MUSE deep-fields: the Ly $\alpha$ luminosity function in the Hubble Deep Field-South at 2.91< z< 6.64},
  author={Drake, Alyssa B and Guiderdoni, Bruno and Blaizot, J{\'e}r{\'e}my and Wisotzki, Lutz and Herenz, Edmund Christian and Garel, Thibault and Richard, Johan and Bacon, Roland and Bina, David and Cantalupo, Sebastiano and others},
  journal={MNRAS},
  volume={471},
  number={1},
  pages={267--278},
  year={2017},
  publisher={Oxford University Press}
}

@article{hoag2019constraining,
  title={Constraining the Neutral Fraction of Hydrogen in the IGM at Redshift 7.5},
  author={Hoag, Austin and Brada{\v{c}}, Maru{\v{s}}a and Huang, K and Mason, Charlotte and Treu, Tommaso and Schmidt, Kasper B and Trenti, Michele and Strait, Victoria and Lemaux, Brian C and Finney, EQ and others},
  journal={ApJ},
  volume={878},
  number={1},
  pages={12},
  year={2019},
  publisher={IOP Publishing}
}

@article{shibuya2019morphologies,
  title={Morphologies of~ 190,000 Galaxies at z= 0--10 Revealed with HST Legacy Data. III. Continuum Profile and Size Evolution of Ly$\alpha$ Emitters},
  author={Shibuya, Takatoshi and Ouchi, Masami and Harikane, Yuichi and Nakajima, Kimihiko},
  journal={ApJ},
  volume={871},
  number={2},
  pages={164},
  year={2019},
  publisher={IOP Publishing}
}

@article{trott2020deep,
  title={Deep multiredshift limits on Epoch of Reionization 21 cm power spectra from four seasons of Murchison Widefield Array observations},
  author={Trott, Cathryn M and Jordan, Christopher H and Midgley, S and Barry, Nichole and Greig, B and Pindor, B and Cook, JH and Sleap, Gregory and Tingay, SJ and Ung, Daniel and others},
  journal={MNRAS},
  volume={493},
  number={4},
  pages={4711--4727},
  year={2020},
  publisher={Oxford University Press}
}

@article{abdurashidova2022first,
  title={First results from hera phase i: Upper limits on the epoch of reionization 21 cm power spectrum},
  author={Abdurashidova, Zara and Aguirre, James E and Alexander, Paul and Ali, Zaki S and Balfour, Yanga and Beardsley, Adam P and Bernardi, Gianni and Billings, Tashalee S and Bowman, Judd D and Bradley, Richard F and others},
  journal={ApJ},
  volume={925},
  number={2},
  pages={221},
  year={2022},
  publisher={IOP Publishing}
}

@article{davies2025efficient,
  title={Efficient simulation of discrete galaxy populations and associated radiation fields during the first billion years},
  author={Davies, James E and Mesinger, Andrei and Murray, Steven},
  journal={arXiv preprint arXiv:2504.17254},
  year={2025}
}

@article{munoz2022impact,
  title={The impact of the first galaxies on cosmic dawn and reionization},
  author={Munoz, Julian B and Qin, Yuxiang and Mesinger, Andrei and Murray, Steven G and Greig, Bradley and Mason, Charlotte},
  journal={MNRAS},
  volume={511},
  number={3},
  pages={3657--3681},
  year={2022},
  publisher={Oxford University Press}
}

@article{qin2020tale,
  title={A tale of two sites--I. Inferring the properties of minihalo-hosted galaxies from current observations},
  author={Qin, Yuxiang and Mesinger, Andrei and Park, Jaehong and Greig, Bradley and Mu{\~n}oz, Julian B},
  journal={MNRAS},
  volume={495},
  number={1},
  pages={123--140},
  year={2020},
  publisher={Oxford University Press}
}

@article{ocvirk2016cosmic,
  title={Cosmic Dawn (CoDa): the first radiation-hydrodynamics simulation of reionization and galaxy formation in the Local Universe},
  author={Ocvirk, Pierre and Gillet, Nicolas and Shapiro, Paul R and Aubert, Dominique and Iliev, Ilian T and Teyssier, Romain and Yepes, Gustavo and Choi, Jun-Hwan and Sullivan, David and Knebe, Alexander and others},
  journal={MNRAS},
  volume={463},
  number={2},
  pages={1462--1485},
  year={2016},
  publisher={Oxford University Press}
}

@article{lee2025line,
  title={Line Intensity Mapping Prediction from the Cosmic Dawn (CoDa). III. Simulation for H $\alpha$ from Galaxies and the Intergalactic Medium During the Epoch of Reionization},
  author={Lee, Eugene Hyeonmin and Lee, Joohyun and Shapiro, Paul R and Ocvirk, Pierre and Lewis, Joseph SW and Dawoodbhoy, Taha and Iliev, Ilian T and Conaboy, Luke and Ahn, Kyungjin and Park, Hyunbae and others},
  journal={Research Notes of the AAS},
  volume={9},
  number={4},
  pages={96},
  year={2025},
  publisher={The American Astronomical Society}
}

@article{maity2023efficient,
  title={Efficient exploration of reionization parameters for the upcoming 21 cm observations using a photon-conserving seminumerical model SCRIPT},
  author={Maity, Barun and Choudhury, Tirthankar Roy},
  journal={MNRAS},
  volume={521},
  number={3},
  pages={4140--4155},
  year={2023},
  publisher={Oxford University Press}
}

@article{prelogovic2023exploring,
  title={Exploring the likelihood of the 21-cm power spectrum with simulation-based inference},
  author={Prelogovi{\'c}, David and Mesinger, Andrei},
  journal={MNRAS},
  volume={524},
  number={3},
  pages={4239--4255},
  year={2023},
  publisher={Oxford University Press}
}

@article{zhao2022simulation,
  title={Simulation-based inference of reionization parameters from 3d tomographic 21 cm light-cone images},
  author={Zhao, Xiaosheng and Mao, Yi and Cheng, Cheng and Wandelt, Benjamin D},
  journal={ApJ},
  volume={926},
  number={2},
  pages={151},
  year={2022},
  publisher={IOP Publishing}
}

@incollection{chapman2019foregrounds,
  title={Foregrounds and their mitigation},
  author={Chapman, Emma and Jeli{\'c}, Vibor},
  booktitle={The Cosmic 21-cm Revolution: Charting the first billion years of our universe},
  pages={6--1},
  year={2019},
  publisher={IOP Publishing Bristol, UK}
}

@article{bonaldi2025square,
  title={Square Kilometre Array Science Data Challenge 3a: foreground removal for an EoR experiment},
  author={Bonaldi, A and Hartley, P and Braun, R and Purser, S and Acharya, A and Ahn, K and Resco, M Aparicio and Bait, O and Bianco, M and Chakraborty, A and others},
  journal={arXiv preprint arXiv:2503.11740},
  year={2025}
}

@article{ShimabukuroSemelin,
    author = {Shimabukuro, Hayato and Semelin, Benoit},
    title = {Analysing the 21 cm signal from the epoch of reionization with artificial neural networks},
    journal = {MNRAS},
    volume = {468},
    number = {4},
    pages = {3869-3877},
    year = {2017},
    month = {03},
    issn = {0035-8711},
    doi = {10.1093/mnras/stx734},
    url = {https://doi.org/10.1093/mnras/stx734},
    eprint = {https://academic.oup.com/mnras/article-pdf/468/4/3869/13944236/stx734.pdf},
}

@article{Jennings2018,
    author = {Jennings, W D and Watkinson, C A and Abdalla, F B and McEwen, J D},
    title = {Evaluating machine learning techniques for predicting power spectra from reionization simulations},
    journal = {MNRAS},
    volume = {483},
    number = {3},
    pages = {2907-2922},
    year = {2018},
    month = {11},
    issn = {0035-8711},
    doi = {10.1093/mnras/sty3168},
    url = {https://doi.org/10.1093/mnras/sty3168},
    eprint = {https://academic.oup.com/mnras/article-pdf/483/3/2907/27212346/sty3168.pdf},
}

@article{Haiman_1997,
doi = {10.1086/303647},
url = {https://dx.doi.org/10.1086/303647},
year = {1997},
month = {feb},
publisher = {},
volume = {476},
number = {2},
pages = {458},
author = {Haiman, Zoltán and Rees, Martin J. and Loeb, Abraham},
title = {Destruction of Molecular Hydrogen during Cosmological Reionization},
journal = {ApJ}
}

@article{Shapiro_2006,
doi = {10.1086/504972},
url = {https://dx.doi.org/10.1086/504972},
year = {2006},
month = {aug},
publisher = {},
volume = {646},
number = {2},
pages = {681},
author = {Shapiro, Paul R. and Ahn, Kyungjin and Alvarez, Marcelo A. and Iliev, Ilian T. and Martel, Hugo and Ryu, Dongsu},
title = {The 21 cm Background from the Cosmic Dark Ages: Minihalos and the Intergalactic Medium before Reionization},
journal = {ApJ}
}

@article{Haiman_2000,
doi = {10.1086/308723},
url = {https://dx.doi.org/10.1086/308723},
year = {2000},
month = {may},
publisher = {},
volume = {534},
number = {1},
pages = {11},
author = {Haiman, Zoltán and Abel, Tom and Rees, Martin J.},
title = {The Radiative Feedback of the First Cosmological
Objects},
journal = {ApJ}
}

@article{Barkana_2002,
doi = {10.1086/342313},
url = {https://dx.doi.org/10.1086/342313},
year = {2002},
month = {oct},
publisher = {},
volume = {578},
number = {1},
pages = {1},
author = {Barkana, Rennan and Loeb, Abraham},
title = {Effective Screening Due to Minihalos during the Epoch of Reionization},
journal = {ApJ}
}

@article{Robertson_2015,
doi = {10.1088/2041-8205/802/2/L19},
url = {https://dx.doi.org/10.1088/2041-8205/802/2/L19},
year = {2015},
month = {apr},
publisher = {The American Astronomical Society},
volume = {802},
number = {2},
pages = {L19},
author = {Robertson, Brant E. and Ellis, Richard S. and Furlanetto, Steven R. and Dunlop, James S.},
title = {COSMIC REIONIZATION AND EARLY STAR-FORMING GALAXIES: A JOINT ANALYSIS OF NEW CONSTRAINTS FROM PLANCK AND THE HUBBLE SPACE TELESCOPE},
journal = {ApJL}
}

@article{SobacchiMesinger2014,
    author = {Sobacchi, Emanuele and Mesinger, Andrei},
    title = {Inhomogeneous recombinations during cosmic reionization},
    journal = {MNRAS},
    volume = {440},
    number = {2},
    pages = {1662-1673},
    year = {2014},
    month = {03},
    issn = {0035-8711},
    doi = {10.1093/mnras/stu377},
    url = {https://doi.org/10.1093/mnras/stu377},
    eprint = {https://academic.oup.com/mnras/article-pdf/440/2/1662/18502076/stu377.pdf},
}

@article{Miralda_Escude_2000,
   title={Reionization of the Inhomogeneous Universe},
   volume={530},
   ISSN={1538-4357},
   url={http://dx.doi.org/10.1086/308330},
   DOI={10.1086/308330},
   number={1},
   journal={ApJ},
   publisher={American Astronomical Society},
   author={Miralda‐Escude, Jordi and Haehnelt, Martin and Rees, Martin J.},
   year={2000},
   month=feb, pages={1–16} }

@ARTICLE{Prochaska2009,
       author = {{Prochaska}, J. Xavier and {Worseck}, Gabor and {O'Meara}, John M.},
        title = "{A Direct Measurement of the Intergalactic Medium Opacity to H I Ionizing Photons}",
      journal = {\apjl},
         year = 2009,
        month = nov,
       volume = {705},
       number = {2},
        pages = {L113-L117},
          doi = {10.1088/0004-637X/705/2/L113},
archivePrefix = {arXiv},
       eprint = {0910.0009},
 primaryClass = {astro-ph.CO},
       adsurl = {https://ui.adsabs.harvard.edu/abs/2009ApJ...705L.113P}
}

@article{Mesinger_2012kSZ,
   title={The kinetic Sunyaev-Zel’dovich signal from inhomogeneous reionization: a parameter space study: kSZ signal from patchy reionization},
   volume={422},
   ISSN={0035-8711},
   url={http://dx.doi.org/10.1111/j.1365-2966.2012.20713.x},
   DOI={10.1111/j.1365-2966.2012.20713.x},
   number={2},
   journal={Monthly Notices of the Royal Astronomical Society},
   publisher={Oxford University Press (OUP)},
   author={Mesinger, Andrei and McQuinn, Matthew and Spergel, David N.},
   year={2012},
   month=mar, pages={1403–1417} }

@article{Barkana_2001,
   title={In the beginning: the first sources of light and the reionization of the universe},
   volume={349},
   ISSN={0370-1573},
   url={http://dx.doi.org/10.1016/S0370-1573(01)00019-9},
   DOI={10.1016/s0370-1573(01)00019-9},
   number={2},
   journal={Physics Reports},
   publisher={Elsevier BV},
   author={Barkana, Rennan and Loeb, Abraham},
   year={2001},
   month=jul, pages={125–238} }

@ARTICLE{Simfast21Santos,
       author = {{Santos}, M.~G. and {Ferramacho}, L. and {Silva}, M.~B. and {Amblard}, A. and {Cooray}, A.},
        title = "{Fast large volume simulations of the 21-cm signal from the reionization and pre-reionization epochs}",
      journal = {\mnras},
         year = 2010,
        month = aug,
       volume = {406},
       number = {4},
        pages = {2421-2432},
          doi = {10.1111/j.1365-2966.2010.16898.x},
archivePrefix = {arXiv},
       eprint = {0911.2219},
 primaryClass = {astro-ph.CO},
       adsurl = {https://ui.adsabs.harvard.edu/abs/2010MNRAS.406.2421S}
}

@misc{GRIZZLYGhara,
       author = {{Ghara}, Raghunath},
        title = "{GRIZZLY: 1D radiative transfer code}",
 howpublished = {Astrophysics Source Code Library, record ascl:2310.012},
         year = 2023,
        month = oct,
          eid = {ascl:2310.012},
       adsurl = {https://ui.adsabs.harvard.edu/abs/2023ascl.soft10012G}
}

@article{Schaeffer_2023,
   title={<scp>beorn</scp>: a fast and flexible framework to simulate the epoch of reionization and cosmic dawn},
   volume={526},
   ISSN={1365-2966},
   url={http://dx.doi.org/10.1093/mnras/stad2937},
   DOI={10.1093/mnras/stad2937},
   number={2},
   journal={Monthly Notices of the Royal Astronomical Society},
   publisher={Oxford University Press (OUP)},
   author={Schaeffer, Timothée and Giri, Sambit K and Schneider, Aurel},
   year={2023},
   month=sep, pages={2942–2959} }

@ARTICLE{Visbal2012,
       author = {{Visbal}, Eli and {Barkana}, Rennan and {Fialkov}, Anastasia and {Tseliakhovich}, Dmitriy and {Hirata}, Christopher M.},
        title = "{The signature of the first stars in atomic hydrogen at redshift 20}",
      journal = {\nat},
         year = 2012,
        month = jul,
       volume = {487},
       number = {7405},
        pages = {70-73},
          doi = {10.1038/nature11177},
archivePrefix = {arXiv},
       eprint = {1201.1005},
 primaryClass = {astro-ph.CO},
       adsurl = {https://ui.adsabs.harvard.edu/abs/2012Natur.487...70V}
}

@article{Fialkov2012,
    author = {Fialkov, Anastasia and Barkana, Rennan and Tseliakhovich, Dmitriy and Hirata, Christopher M.},
    title = {Impact of the relative motion between the dark matter and baryons on the first stars: semi-analytical modelling},
    journal = {Monthly Notices of the Royal Astronomical Society},
    volume = {424},
    number = {2},
    pages = {1335-1345},
    year = {2012},
    month = {08},
    issn = {0035-8711},
    doi = {10.1111/j.1365-2966.2012.21318.x},
    url = {https://doi.org/10.1111/j.1365-2966.2012.21318.x},
    eprint = {https://academic.oup.com/mnras/article-pdf/424/2/1335/3204998/424-2-1335.pdf},
}

@article{greig2017simultaneously,
  title={Simultaneously constraining the astrophysics of reionization and the epoch of heating with 21CMMC},
  author={Greig, Bradley and Mesinger, Andrei},
  journal={Monthly Notices of the Royal Astronomical Society},
  volume={472},
  number={3},
  pages={2651--2669},
  year={2017},
  publisher={Oxford University Press}
}

@ARTICLE{2016MNRAS.457.1864L,
       author = {{Liu}, Adrian and {Parsons}, Aaron R.},
        title = "{Constraining cosmology and ionization history with combined 21 cm power spectrum and global signal measurements}",
      journal = {\mnras},
         year = 2016,
        month = apr,
       volume = {457},
       number = {2},
        pages = {1864-1877},
          doi = {10.1093/mnras/stw071},
archivePrefix = {arXiv},
       eprint = {1510.08815},
 primaryClass = {astro-ph.CO},
       adsurl = {https://ui.adsabs.harvard.edu/abs/2016MNRAS.457.1864L}
}

@misc{germain2015mademaskedautoencoderdistribution,
      title={MADE: Masked Autoencoder for Distribution Estimation}, 
      author={Mathieu Germain and Karol Gregor and Iain Murray and Hugo Larochelle},
      year={2015},
      eprint={1502.03509},
      archivePrefix={arXiv},
      primaryClass={cs.LG},
      url={https://arxiv.org/abs/1502.03509}, 
}

@misc{papamakarios2018maskedautoregressiveflowdensity,
      title={Masked Autoregressive Flow for Density Estimation}, 
      author={George Papamakarios and Theo Pavlakou and Iain Murray},
      year={2018},
      eprint={1705.07057},
      archivePrefix={arXiv},
      primaryClass={stat.ML},
      url={https://arxiv.org/abs/1705.07057}, 
}

@misc{zhao2024simulationbasedinferencereionizationparameters,
      title={Simulation-based Inference of Reionization Parameters from 3D Tomographic 21 cm Light-cone Images -- II: Application of Solid Harmonic Wavelet Scattering Transform}, 
      author={Xiaosheng Zhao and Yi Mao and Shifan Zuo and Benjamin D. Wandelt},
      year={2024},
      eprint={2310.17602},
      archivePrefix={arXiv},
      primaryClass={astro-ph.IM},
      url={https://arxiv.org/abs/2310.17602}, 
}

@ARTICLE{2024mertensforegrounds,
       author = {{Mertens}, Florent G. and {Bobin}, J{\'e}r{\^o}me and {Carucci}, Isabella P.},
        title = "{Retrieving the 21-cm signal from the Epoch of Reionization with learnt Gaussian process kernels}",
      journal = {\mnras},
         year = 2024,
        month = jan,
       volume = {527},
       number = {2},
        pages = {3517-3531},
          doi = {10.1093/mnras/stad3430},
archivePrefix = {arXiv},
       eprint = {2307.13545},
 primaryClass = {astro-ph.CO},
       adsurl = {https://ui.adsabs.harvard.edu/abs/2024MNRAS.527.3517M}
}

@ARTICLE{Hyvarinen,

  author={Hyvarinen, A.},

  journal={IEEE Transactions on Neural Networks}, 

  title={Fast and robust fixed-point algorithms for independent component analysis}, 

  year={1999},

  volume={10},

  number={3},

  pages={626-634},

  doi={10.1109/72.761722}}

@ARTICLE{2021MNRAS.500.2264H,
       author = {{Hothi}, Ian and {Chapman}, Emma and {Pritchard}, Jonathan R. and {Mertens}, F.~G. and {Koopmans}, L.~V.~E. and {Ciardi}, B. and {Gehlot}, B.~K. and {Ghara}, R. and {Ghosh}, A. and {Giri}, S.~K. and {Iliev}, I.~T. and {Jeli{\'c}}, V. and {Zaroubi}, S.},
        title = "{Comparing foreground removal techniques for recovery of the LOFAR-EoR 21 cm power spectrum}",
      journal = {\mnras},
         year = 2021,
        month = jan,
       volume = {500},
       number = {2},
        pages = {2264-2277},
          doi = {10.1093/mnras/staa3446},
archivePrefix = {arXiv},
       eprint = {2011.01284},
 primaryClass = {astro-ph.CO},
       adsurl = {https://ui.adsabs.harvard.edu/abs/2021MNRAS.500.2264H}
}

@ARTICLE{2018MNRAS.478.3640M,
       author = {{Mertens}, F.~G. and {Ghosh}, A. and {Koopmans}, L.~V.~E.},
        title = "{Statistical 21-cm signal separation via Gaussian Process Regression analysis}",
      journal = {\mnras},
         year = 2018,
        month = aug,
       volume = {478},
       number = {3},
        pages = {3640-3652},
          doi = {10.1093/mnras/sty1207},
archivePrefix = {arXiv},
       eprint = {1711.10834},
 primaryClass = {astro-ph.CO},
       adsurl = {https://ui.adsabs.harvard.edu/abs/2018MNRAS.478.3640M}
}

@ARTICLE{2012MNRAS.423.2518C,
       author = {{Chapman}, Emma and {Abdalla}, Filipe B. and {Harker}, Geraint and {Jeli{\'c}}, Vibor and {Labropoulos}, Panagiotis and {Zaroubi}, Saleem and {Brentjens}, Michiel A. and {de Bruyn}, A.~G. and {Koopmans}, L.~V.~E.},
        title = "{Foreground removal using FASTICA: a showcase of LOFAR-EoR}",
      journal = {\mnras},
         year = 2012,
        month = jul,
       volume = {423},
       number = {3},
        pages = {2518-2532},
          doi = {10.1111/j.1365-2966.2012.21065.x},
archivePrefix = {arXiv},
       eprint = {1201.2190},
 primaryClass = {astro-ph.CO},
       adsurl = {https://ui.adsabs.harvard.edu/abs/2012MNRAS.423.2518C}
}

%%%%%%%%%%%%%%%%%%%%%%%%%%%%%%%%%%%%%%%%%%%%%%%%%%

%%%%%%%%%%%%%%%%% APPENDICES %%%%%%%%%%%%%%%%%%%%%

\appendix

\section{Ionization History Corner Plots} \label{corner}

\begin{figure*}\label{ion_corner}
	\includegraphics[width=\textwidth]{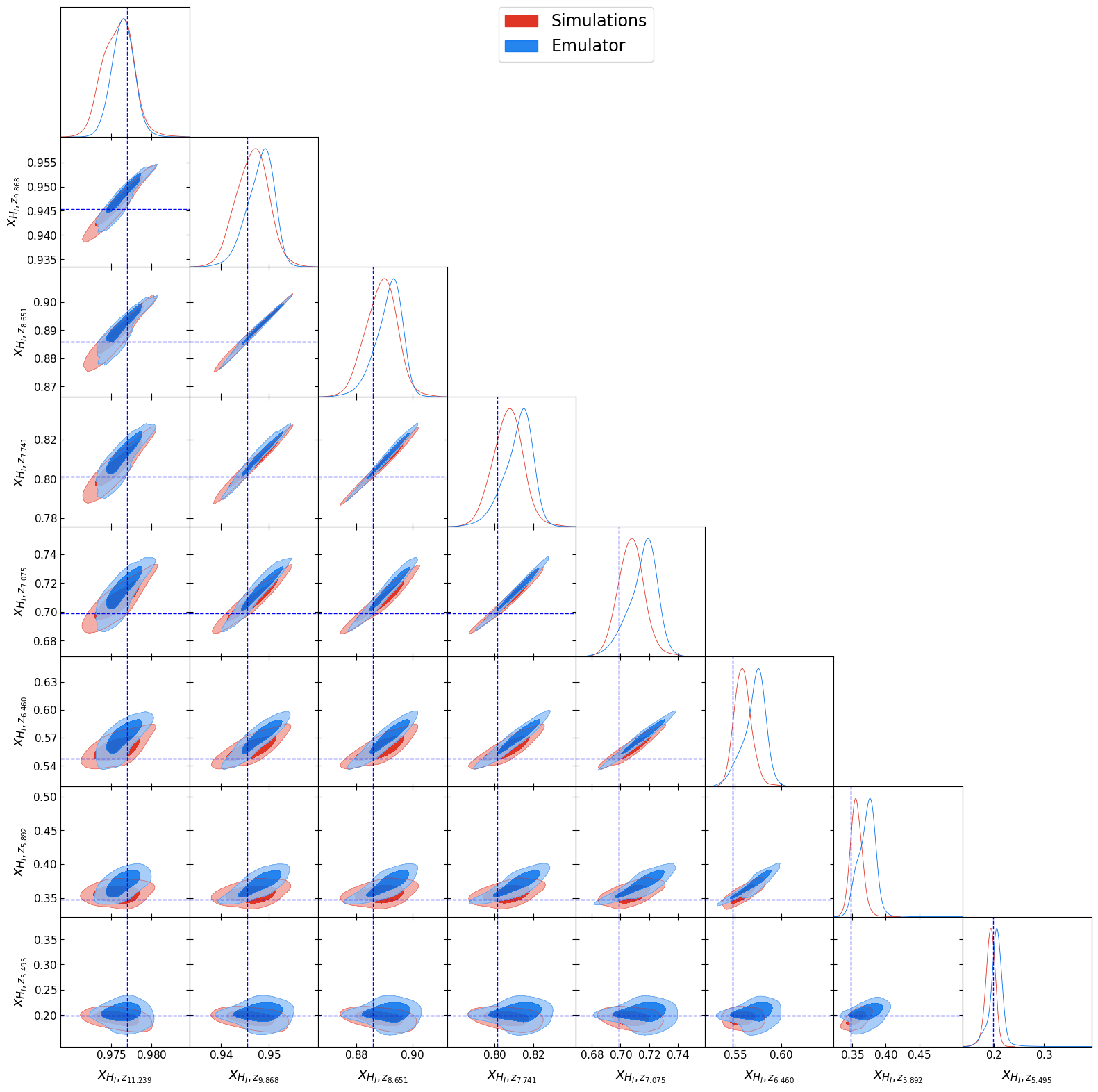}
    \caption{Corner plots for the neutral fraction at eight redshift slices. The red contours are the one and two sigma contours obtained with the pipeline trained on the simulations, and the blue is with emulated data as the training set. Each contour is generated with 10,000 posterior samples inferred from a noisy 2DPS inference. The ground truth ionization history is given by the dashed lines.} 
    \label{fig:ion_corner_plots} 
\end{figure*}

We explored two methods to obtain posterior distributions on the neutral fraction over cosmic time. Constraints on the astrophysical parameters were obtained with simulation based inference. We sample from the obtained posterior distribution. The ionization history at each parameter sample was emulated, and these samples form the posterior for the neutral fraction. The obtained posteriors for an example inference on the two methods can be seen in Fig.  \ref{fig:ion_corner_plots}.

\section{Architectures}\label{architectures}

\begin{figure}

	\includegraphics[width=\columnwidth]{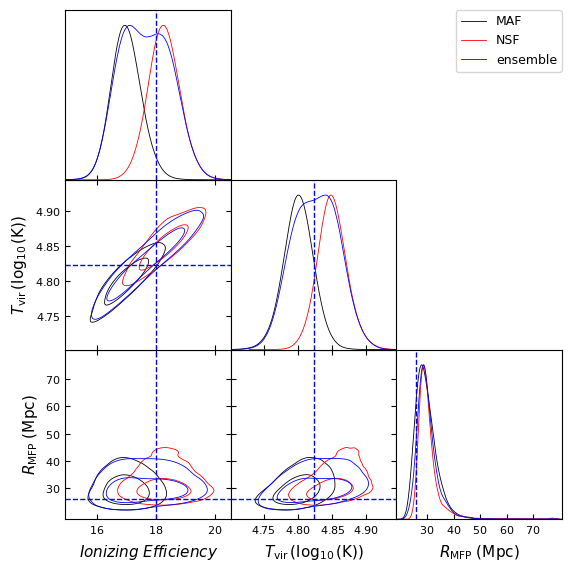}
    \caption{Comparison of simulation based inference architectures trained on noisy 21cmFAST 2DPS simulations. The architectures correspond to those detailed in Table \ref{tab:SBI_arch}, in the same order. The posteriors show the one and two sigma contours for 10,000 posterior samples. The ground truth used to simulate the observation is given by the dashed lines.}
    \label{fig:architectures} 
\end{figure}

In order to select the best architecture for the neural density estimation, 5 architectures were first independently trained, as described in \ref{sec:inference}. The final architectures selected are detailed in Table \ref{tab:SBI_arch}. 
The two neural architectures show non-negligible differences in inferred constraints. These differences can be seen for an example inference from the pipeline trained on simulations in Fig.  \ref{fig:architectures}. 
The uncertainty due to the variance of different neural density techniques can be mitigated for by averaging the samples from each architecture. The averaging is weighted by the loss of the validation set.

Many architectures were tested, and evaluated by their SBC performance. The motivation behind this is that different architectures may be able to capture different features with varying levels of accuracy. It is not evident which specific architecture would best capture the properties of our data. Hence, we investigate different architectures. Since we only included networks in the ensemble which improved the ensemble's SBC constraints, a range of architectures were tried without risk. NSFs and MAFs are both normalizing flows which can be more flexible when capturing non-Gaussian features, as they can transform into more complex distributions. MDNs alternatively, are comprised as a combination of Gaussian components, these can be faster to train but are less flexible. MADEs are an autoencoder which directly model the distribution, MAFs stack MADEs, in order to transform a base distribution \citep{papamakarios2019neural}. \cite{papamakarios2018maskedautoregressiveflowdensity} find that the performance of MAFs vs MADEs can be dataset dependent, hence we tested both architectures.  The two final architectures were chosen as the SBC of the combined ensemble performs better than the SBC of any individual component. The pipeline does not display a clear preference towards either of these architectures, as seen in the performance metrics below which also demonstrate how the ensemble improves performance.

We calculated the bias for each astrophysical parameter for the two different architectures according to 
\begin{equation}
    bias = \frac{\Bar{\theta} - \theta_{true}}{1 \sigma}
\end{equation}

where $\Bar{\theta}$ is the mean value of the samples for each architecture and $\sigma$ is the standard deviation. The results of this can be seen in Table \ref{tab:bias_table}. As Table B1 indicates, across the three parameters, the MAF has the lowest individual bias. The bias is then further reduced by using the ensemble network.

We also calculated the deviation from perfectly flat SBC histograms for each architecture. We did this by calculating the area between each bin and the flat perfect calibration line. Each histogram is normalized to one; so a value of zero is perfect calibration, and a value of one indicates all the samples in one bin. The results of this can be seen in Table \ref{tab:SBC_table}. The results of the SBC do not align perfectly with the bias metrics, as we see the MAF performing with the larger deviation from flatness in some parameters. We therefore average over both architectures as each parameter and performance metric displays a preference for different architectures. This averaging is weighted by the value of the validation loss, favoring higher performing models. As Table \ref{tab:SBC_table} indicates, the SBC calibration is improved when using the ensemble architecture, justifying the decision to use a mixture of both networks.

\begin{table}
	\centering
	\caption{Bias of each neural network architecture for each astrophysical parameter. The bias was calculated using 10,000 posterior samples. }
	\label{tab:bias_table}
	\begin{tabular}{lcccr} 
		\hline
		Architecture & $\zeta$ & $\mathrm{T_{vir}}$ & $\mathrm{R_{MFP}}$&sum \\
		\hline
		MAF & 0.025 & 0.049 & 0.009 & 0.164\\
		NSF & 0.061 & 0.041 & 0.160 & 0.262\\
        Ensemble & 0.022 & 0.008 & 0.080 & 0.110\\
        \end{tabular}

\end{table}

\begin{table}
	\centering
	\caption{Deviation from flatness of the SBC for both neural network architectures for each astrophysical parameter. The SBC was calculated using 10,000 posterior samples, and the area for each histogram normalized to one. Zero represents perfect calibration, one would be complete deviation (all parameters in one bin).}
	\label{tab:SBC_table}
	\begin{tabular}{lcccr} 
		\hline
		Architecture & $\zeta$ & $\mathrm{T_{vir}}$ & $\mathrm{R_{MFP}}$ & sum \\
		\hline
		MAF & 0.062 & 0.063 & 0.075 & 0.200\\
		NSF & 0.042 & 0.043 & 0.118 & 0.203\\
        Ensemble & 0.061 & 0.043 & 0.080 & 0.184 \\
        \end{tabular}

\end{table}

%%%%%%%%%%%%%%%%%%%%%%%%%%%%%%%%%%%%%%%%%%%%%%%%%%

% Don't change these lines
\bsp	% typesetting comment
\label{lastpage}
\end{document}